\documentclass[usenatbib]{mnras}

\usepackage{newtxtext,newtxmath}

\usepackage[T1]{fontenc}
\usepackage{ae,aecompl}

\usepackage{graphicx}	
\usepackage{amsmath}	
\usepackage{amssymb}	

\title[Dynamical mass estimation]{Constraining cluster masses from the stacked phase space distribution at large radii}

\author[A. Hamabata et al.]{
Akinari Hamabata,$^{1}$\thanks{E-mail:hamabata@utap.phys.s.u-tokyo.ac.jp}
Masamune Oguri,$^{1,2,3}$
and Takahiro Nishimichi$^{2}$
\\
$^{1}$Department of Physics, The University of Tokyo, 7-3-1 Hongo, Bunkyo-ku, Tokyo 113-0033, Japan\\
$^{2}$Kavli Institute for the Physics and Mathematics of the Universe, University of Tokyo, Kashiwa, Chiba 277-8583, Japan\\
$^{3}$Research Center for the Early Universe, The University of Tokyo, 7-3-1 Hongo, Bunkyo-ku, Tokyo 113-0033, Japan
}

\date{Accepted XXX. Received YYY; in original form ZZZ}

\pubyear{2018}

\begin{document}
\label{firstpage}
\pagerange{\pageref{firstpage}--\pageref{lastpage}}
\maketitle

\begin{abstract}
Velocity dispersions have been employed as a method to measure masses of clusters.
To complement this conventional method, we explore the possibility of constraining cluster masses from the stacked phase space distribution of galaxies at larger radii, where infall velocities are expected to have a sensitivity to cluster masses.
First, we construct a two component model of the three-dimensional phase space distribution of haloes surrounding clusters up to 50 $h^{-1}$Mpc from cluster centres based on $N$-body simulations.
We find that the three-dimensional phase space distribution shows a clear cluster mass dependence up to the largest scale examined.
We then calculate the probability distribution function of pairwise line-of-sight velocities between clusters and haloes by projecting the three-dimensional phase space distribution along the line-of-sight with the effect of the Hubble flow.
We find that this projected phase space distribution, which can directly be compared with observations, shows a complex mass dependence due to the interplay between infall velocities and the Hubble flow.
Using this model, we estimate the accuracy of dynamical mass measurements from the projected phase space distribution at the transverse distance from cluster centres larger than $2h^{-1}$Mpc.
We estimate that, by using $1.5\times 10^5$ spectroscopic galaxies, we can constrain the mean cluster masses with an accuracy of 14.5\% if we fully take account of the systematic error coming from the inaccuracy of our model.
This can be improved down to 5.7\% by improving the accuracy of the model.
\end{abstract}




\section{Introduction}
\label{SI}
In the framework of hierarchical structure formation, galaxy clusters, which are the biggest self-gravitating system in the Universe, represent the current end point of the evolution of primordial density fluctuations.
Thus, we can extract information on the initial density perturbation, the growth of structure, and cosmological parameters from observations of galaxy clusters.
For instance, we can extract the matter density ($\Omega_{\rm{m}}$) and the amplitude of the density flctuation ($\sigma_{8}$) from the abundance of galaxy clusters (e.g., \citealt{Rozo2010}).
The uncertainty of estimating cluster masses, which are necessary to compare observations with theory involving dark matter, is one of the dominant sources of the uncertainty in such cosmological analyses.
To constrain cosmological parameters accurately, we need to estimate masses of galaxy clusters accurately, which is difficult because masses of clusters are dominated by invisible dark matter.

There are several methods to estimate masses of galaxy clusters, including gravitational lensing (e.g., \citealt{Schneider1992}; \citealt{Umetsu2011}; \citealt{Oguri2012}; \citealt{Newman2013}; \citealt{Okabe2016}), X-ray observations (e.g., \citealt{Sarazin1988}; \citealt{Vikhlinin}), and the Sunyaev-Zel'dovich (SZ) effect (e.g., \citealt{SZ}; \citealt{Arnaud2010}; \citealt{Plank2014}).
In addition, we can estimate masses of galaxy clusters by using the relative motion of galaxies surrounding galaxy clusters.
Because the gravitational potential of galaxy clusters affects the motions of surrounding galaxies, such motions have information on cluster masses.
In this paper, we refer to the mass estimated by motions of galaxies around clusters as the dynamical mass.
It is of great importance to compare cluster masses derived by these different methods in order to understand systematic errors inherent to the individual methods.
Different methods have different systematic errors, which can be inferred and hopefully corrected for by cross-checking the results of the individual methods.

Since the pioneering work by \cite{Smith1936} who estimated the mass of the Virgo cluster by using motions of galaxies, many papers have studied dynamical masses of galaxy clusters (e.g., \citealt{Busha+2005}; \citealt{Rozo2015}; \citealt{Farahi2016}).
However, there is a room for improvement in several ways.
For example, to estimate dynamical masses accurately, we have to understand the dynamical state of galaxies around galaxy clusters, taking account of motions of galaxies inside clusters as well as those of infalling galaxies.
The caustic model is a method to extract the infall sequence to galaxy clusters based on a spherical collapse model (\citealt{Diaferio1997}), which is extended to an expanding Universe in \cite{Stark2016}.
However, there is an ambiguity to define the caustic surface in observations due to the projection effect, which has to be calibrated against $N$-body simulations.
In most of the previous studies, motions of galaxies within the transverse distance of $\sim 1~ \rm{Mpc} $ from centres of galaxy clusters are used to derive dynamical masses.
However, these studies mainly focused on velocity dispersions and did not consider the full distribution of line-of-sight velocities, nor their radial dependence.
Furthermore, it has been known that cosmological $N$-body simulations, which are very often used to connect velocity measurements around clusters with cluster masses, contain a lot of uncertainty because they do not include baryonic effects, such as radiative cooling, gas pressure, and feedbacks.
Such baryonic effects may change the mass profiles and dynamics of galaxies around galaxy clusters, because the efficiency of tidal stripping and dynamical friction can be significantly affected by baryonic effects (e.g., \citealt{Suto2017}; \citealt{Peirani2017}).

One important application of dynamical mass measurements is to test General Relativity by comparing cluster masses estimated by gravitational lensing effect with dynamical masses (e.g., \citealt{Schmidt2010}; \citealt{Lam2012}; \citealt{LAM2013}).
Because the dynamical mass and the lensing mass contain different information on the spacetime metric, we can test General Relativity by comparing these two masses.
However, in some classes of modified gravity, such as $f(R)$ gravity (e.g., \citealt{Nojiri+2011}), the deviation from General Relativity is suppressed at the regions close to cluster centres where the gravity is relatively strong.
Hence, to test General Relativity efficiency, it is important to study the outer part of cluster mass profiles (\citealt{Lombriser2014}; \citealt{Terukina2015}).

In this paper, we study the phase space distribution of galaxies around galaxy clusters on large separations well beyond the conventional radii used for dynamical measurements, up to several tens of Mpc, by using $N$-body simulations, and propose to use the stacked phase space distribution at these large radii to constrain cluster masses.
For this purpose, we construct a new analytical model of the phase space distribution around clusters.
While our model of the phase space distribution is similar to those shown in \cite{Zu2013} and \cite{LAM2013} in several ways, we explicitly consider the dependence on the cluster mass so that it can be used to estimate the dynamical mass.
After we construct a model of the full three-dimensional phase space distribution, we take account the impact of the Hubble flow, which is indistinguishable from peculiar velocities of galaxies on an individual basis.
We discuss how we can treat this effect in projecting the phase space model to finally obtain the projected pairwise line-of-sight velocity distribution between galaxies and clusters.

In this paper, we demonstrate how dynamical masses can be estimated by the stacked phase space distribution at large transverse distances beyond $\sim 2~h^{-1}  {\rm{Mpc}}$, which is highly complementary to traditional methods to estimate dynamical masses from motions of galaxies within the transverse distance of $\sim 1~ h^{-1} {\rm{Mpc}} $ from cluster centres.
A possible advantage of using the outer phase space distribution is that it is expected to be much less sensitive to baryonic effects compared with the phase space distribution within $\sim 1 ~ h^{-1}$Mpc as discussed above.
This suggests that the observed phase space distribution may be compared with $N$-body simulations more directly at such large distances from cluster centres.

This paper is organised as follows.
In Section \ref{S2.0}, we show our model of the three-dimensional phase space distribution around clusters.
In Section \ref{S3.0}, we project the three-dimensional distribution along the line-of-sight with the effect of the Hubble flow.
In Section \ref{S4.0}, we discuss how the dynamical mass is measured using our model.
Finally, we conclude in Section \ref{SD}.

\section{Three-Dimensional Phase Space Distribution}
\label{S2.0}

\subsection{Simulations}

We perform four random realizations of cosmological $N$-body simulations.
Each simulation is performed with a TreePM code $\tt{Gadget}$-$2$ \citep{Springel2005}, which runs from $z = 99$ to $0$ in a box of comoving $360 ~ h^{-1} \rm{Mpc} $ on a side with the periodic boundary condition.
The number of dark matter particles is $1024^{3}$, corresponding to the mass of each particle of $m_{p} = 3.4 \times 10^{9}  h^{-1} M_{\odot} $.
The gravitational softening length is fixed at comoving $20  ~ h^{-1} \rm{kpc} $.
The initial condition is generated by a code based on the second order Lagrangian perturbation theory to compute the displacement from the regular lattice preinitial configuration developed in \citet{Nishimichi2009} and parallelised in \citet{Valageas2011}.
The transfer function at $z = 99$ is computed by the linear Boltzman code $\tt{CAMB}$ \citep{Lewis2000}.
We adopt $\Omega_{M,0} = 0.279$, $\Omega_{\Lambda,0} = 0.721$, $h = 0.7$, $n_{s} = 0.972$, and $\sigma_{8} = 0.821$ following the WMAP nine year result \citep{Hinshaw2013}.
To identify haloes\footnote{{\tt{Rockstar}} identifies both haloes and subhaloes from $N$-body simulations. In this paper, we do not distinguish haloes and subhaloes, and collectively call them haloes.} in our $N$-body simulations, we use six-dimensional friend of friend (FoF) algorithm implemented in $\tt{Rockstar}$ \citep{Behroozi2013B}.

\subsection{Stacked Phase Space Distribution}

\begin{figure}
    \includegraphics[width=\columnwidth]{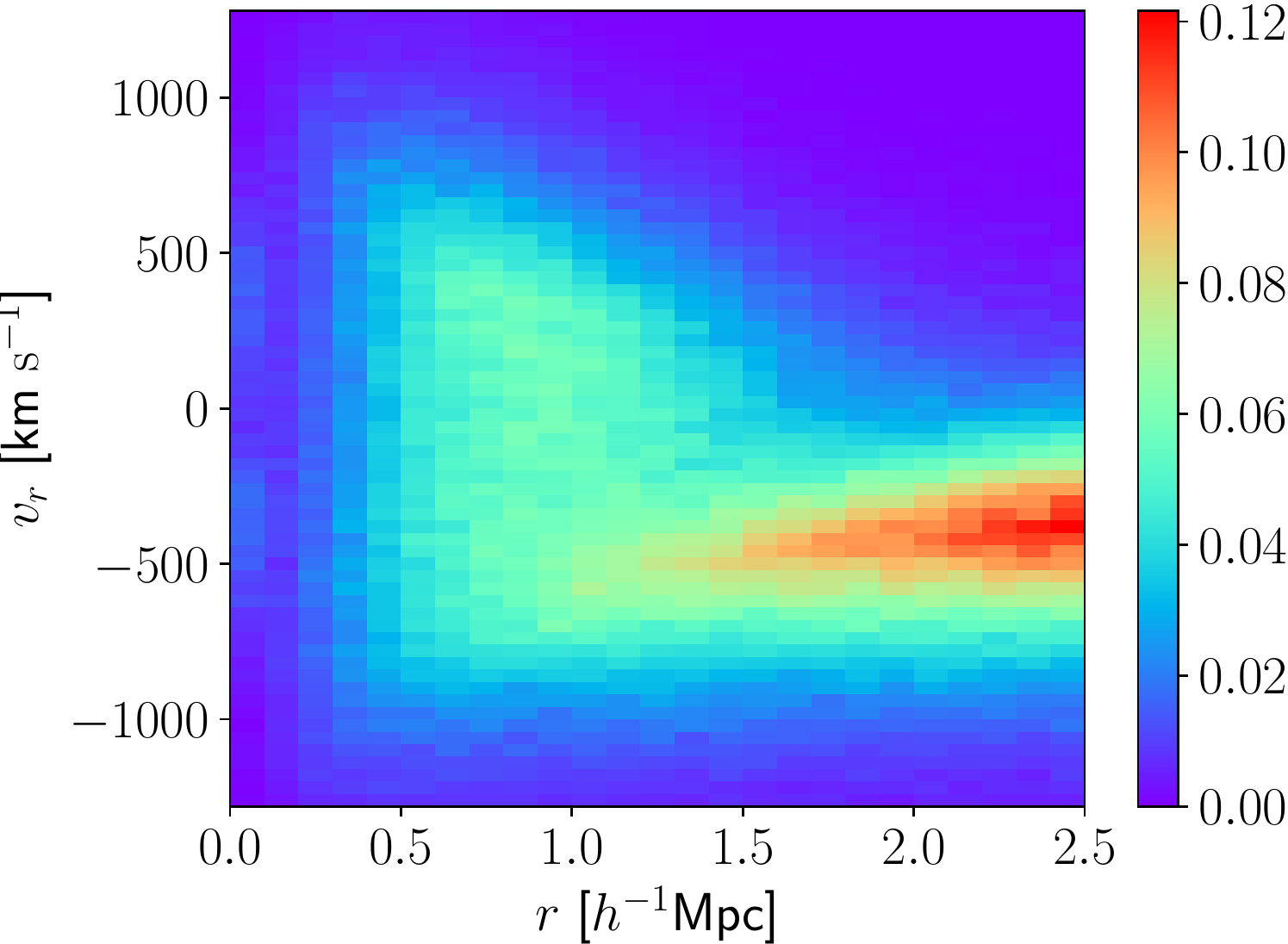}
    \caption{The stacked three-dimensional phase space distribution of dark matter haloes at $z=0$, obtained from our $N$-body simulations.
Only haloes with $M_{\rm{halo}} > 1 \times 10^{11}  ~ h^{-1} M_\odot$ are used.
The mass range of clusters is $0.560\times 10^{14} ~ h^{-1} M_\odot < M_{\rm{cl}} < 3.679 \times 10^{14} ~ h^{-1} M_\odot$.
The vertical axis is the radial velocity of haloes, which is defined such that positive $v_{r}$ corresponds to outward motions.
The horizontal axis is the radius from the centres of galaxy clusters.
The colour scale shows the number density of haloes in the phase space, $\log{f(v_{r})}$, which is defined as the number density of haloes per each galaxy cluster with bin sizes of 40 km $\rm{s^{-1}}$ for $v_{r}$ bin and 0.2 $h^{-1} M_\odot$ for $r$ bin.
Note that we sum up all haloes within the spherical shell.}
    \label{Fig1}
\end{figure}

We use these simulations to obtain the phase space distribution of dark matter haloes around galaxy clusters.
Because we are interested in average features of dynamics of dark matter haloes, we stack simulated haloes around a large number of galaxy clusters to derive accurate three-dimensional phase space distributions.
We show an example of stacked phase space distribution from $N$-body simulations in Fig, \ref{Fig1}.

We use haloes with masses $M_{\rm halo}> 1 \times 10^{11} h^{-1} M_\odot $ in this paper.
Note that we define  $M_{\rm halo}$ as the virial mass of halo.
The virial mass is the mass within $r_{\rm vir}$, which is the radius within which the average density is $\Delta_{\rm{vir}}$ times the background density at $z = 0$, which is calculated following \cite{Bryan1998}.
We also define cluster mass ($M_{\rm{cl}}$) as the virial mass of galaxy cluster.
We define galaxy clusters as haloes with masses $0.560 \times 10^{14} h^{-1} M_\odot < M_{\rm cl} <  3.679 \times 10^{14} h^{-1} M_\odot $.
To mimic observations, we remove galaxy clusters from our analysis if there are any other clusters with larger masses within $1  ~ h^{-1} \rm{Mpc} $ from those clusters.
To analyse the cluster mass dependence of the phase space distribution, we divide galaxy clusters into 24 log-equal mass bins.
We regard the geometric mean mass of the lower and upper limits of each mass bin as the mass of the bin.
Note that galaxy clusters can become haloes when we focus on other galaxy clusters.
We use only snapshots at $z = 0$ in this paper for simplicity.

To estimate cluster masses from the line-of-sight velocity ($v_{\rm{los}}$) histogram, we first construct a model of the three-dimensional phase space distribution, and then project it along the line-of-sight to obtain the histogram of $v_{\rm los}$.
The three-dimensional phase space distribution shown in Fig. \ref{Fig1} clearly indicates that the phase space distribution is quite complicated.
In what follows, we construct a two component model to take account of the complexity of the phase space distribution.

\subsection{Fitting with a Two Component Model}
\label{S2.2}
We construct a model of the three-dimensional phase space distribution of haloes surrounding galaxy clusters based on the stacked phase space distributions obtained from the $N$-body simulations.
Because we stack a large number of galaxy clusters without aligning their orientations, the spherical asymmetry of the phase space distributions for individual clusters should be averaged out.
Hence, throughout the paper we assume a spherically symmetric phase space distribution.
\begin{figure}
    \includegraphics[width=\columnwidth]{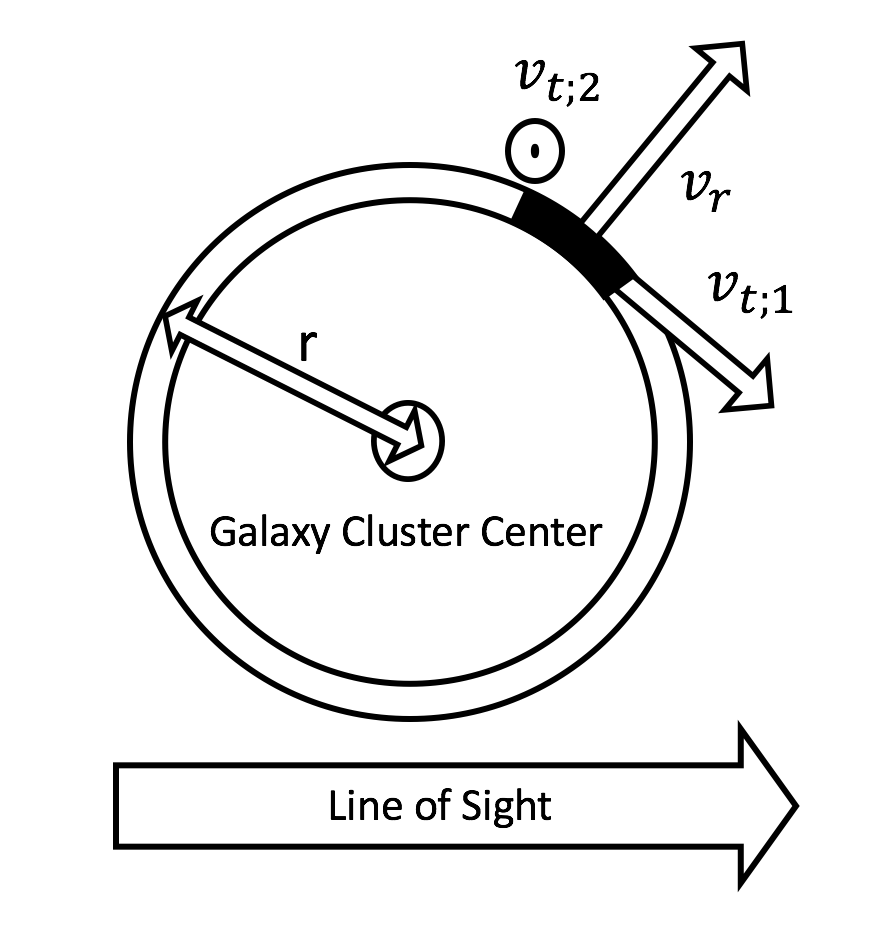}
    \caption{Schematic picture of three velocity components.}
    \label{P1}
\end{figure}

We divide a velocity of a halo into three orthogonal components, the radial velocity ($v_{r}$) and two tangential velocities ($v_{t;1} , \ v_{t;2} $) as shown in Fig. \ref{P1}.
At this point we only consider pairwise peculiar velocities between clusters and haloes and do not consider the Hubble flow.
Since one of the two components of the tangential velocities do not contribute to $v_{\rm{los}}$, we ignore $v_{t;2} $ in this paper, and denote $v_{t;1}$ as $v_{t}$.

Under the assumption of the spherically symmetric phase space distribution, we can describe the probability distribution function (PDF) of the phase space as
 \begin{equation}
 \label{eq2-1}
 p_{v} = p_{v}(v_{r} , v_{t} ,r) \ .
\end{equation}
We then assume that the PDF of the phase space distribution can be divided into two components, the infall component and the splashback component.
The infall component corresponds to haloes that are now falling into galaxy clusters, and the splashback component corresponds to haloes that are on their first (or more) orbit after falling into galaxy clusters.
Such two component model is also proposed in \citet{Zu2013}, although they consider a virial component for which the average value of the radial distribution is fixed to zero.
In contrast, in our splashback component we allow a non-zero average radial velocity.
In the two component model, equation (\ref{eq2-1}) is described as
 \begin{equation}
 \label{eq2-2}
p_{v}(v_{r} , v_{t} ,r) = ( 1 - \alpha )  p_{\rm{infall}} (v_{r} , v_{t} ,r)  + \alpha    p_{\rm{SB}} (v_{r} , v_{t} ,r) ,
\end{equation}
where both $p_{\rm{infall}}$ and $p_{\rm{SB}}$ are properly normalised, and thus $\alpha  $ denotes the fraction of the splashback component at given $r$.
For simplicity, we also assume that there is no correlation between radial and tangential velocities in both of two components.
With this assumption, we can rewrite equation (\ref{eq2-2}) as 
 \begin{equation}
 \label{eq2-3}
 \begin{split}
p_{v}(v_{r} , v_{t} ,r) = \ & (1 - \alpha )  p_{v_{r},\rm{infall}} (v_{r} ,r)  p_{v_{t},\rm{infall}} (v_{t} ,r) \\
& + \alpha   p_{v_{r},\rm{SB}} (v_{r}  ,r)  p_{v_{t},\rm{SB}} (v_{t}  ,r)\ .
\end{split}
\end{equation}
We discuss each component one by one in what follows.
\subsubsection{Radial Velocity Distribution}
We start with the distribution of the radial velocities.
We find from the simulations that the radial velocity distributions at large radii, where the distributions are dominated by the infall component, have non-negligible skewness and kurtosis, as was already shown in \citet{scoccimarro2004}. To incorporate the skewness and kurtosis, we adopt the Johnson's SU-distribution \citep{Jhonson1949} as the model function for the radial velocity distribution of the infall component,
 \begin{equation}
 \label{eq2-4}
  \begin{split}
p_{v_{r},\rm{infall}} & (v_{r} , r) =\  SU (v_{r};\delta_{r},\lambda_{r},\gamma_{r}, \xi_{r} ) \\
 =\ & \frac{\delta_{r} }{\lambda_{r} \sqrt{2 \pi} \sqrt{ \{ z(v_{r}) \}^2 + 1}} \exp{ \left[ - \frac{1}{2}   \left( \gamma_{r} + \delta_{r} \sinh^{-1}{z(v_{r})} \right )    \right]},
 \end{split}
\end{equation}
where
 \begin{equation}
 \label{eq2-5}
z(v_{r}) = \frac{v_{r} - \xi_{r}}{ \lambda_{r}}.
\end{equation}
The Johnson's SU-distribution has four free parameters, and ths is sufficiently flexible to model the phase space distributions from $N$-body simulations including the skewness and kurtosis.
We note that these four parameters, $\delta_{r},\lambda_{r},\gamma_{r}$, and $\xi_{r}$ are functions of the radius $r$.

As for the splashback component, on the other hand, we adopt the Gaussian distribution
 \begin{equation}
 \label{eq2-6}
   \begin{split}
p_{v_{r},\rm{SB}} (v_{r} , r) =\  & G (v_{r} ; \mu_{r}  , \sigma_{r}^{2} ) \\
=\  & \frac{1}{\sqrt{2 \pi \sigma_{r}^2 }} \exp{ \left \{ - \frac{ ( v_{r} - \mu_{r}  ) ^ 2}{ 2 \sigma_{r}^2 } \right \} } \ ,
\end{split}
\end{equation}
since we do not find a clear evidence for higher order cumulants.
Again, we note that the free parameters, $\mu_{r}$ and $\sigma_{r}^{2}$, are determined as functions of $r$.

The model function for the radial velocity distribution is
 \begin{equation}
 \begin{split}
 \label{eq2-7}
p_{v_{r}} (v_{r} , r) =  & \ \int d v_{t} (1 - \alpha )  p_{v_{r},\rm{infall}} (v_{r} ,r)  p_{v_{t},\rm{infall}} (v_{t} ,r) \\
& + \alpha   p_{v_{r},\rm{SB}} (v_{r}  ,r)  p_{v_{t},\rm{SB}} (v_{t}  ,r) \\
= & \ (1 - \alpha) p_{v_{r},\rm{infall}} + \alpha p_{v_{r},\rm{SB}}\ .
\end{split}
\end{equation}
For a given radial and mass bin, we fit the histogram of radial velocities obtained from our $N$-body simulations with equation (\ref{eq2-7}) that contains seven parameters.

\begin{figure}
 \begin{minipage}{0.45\textwidth}
    \includegraphics[width=1.0\textwidth]{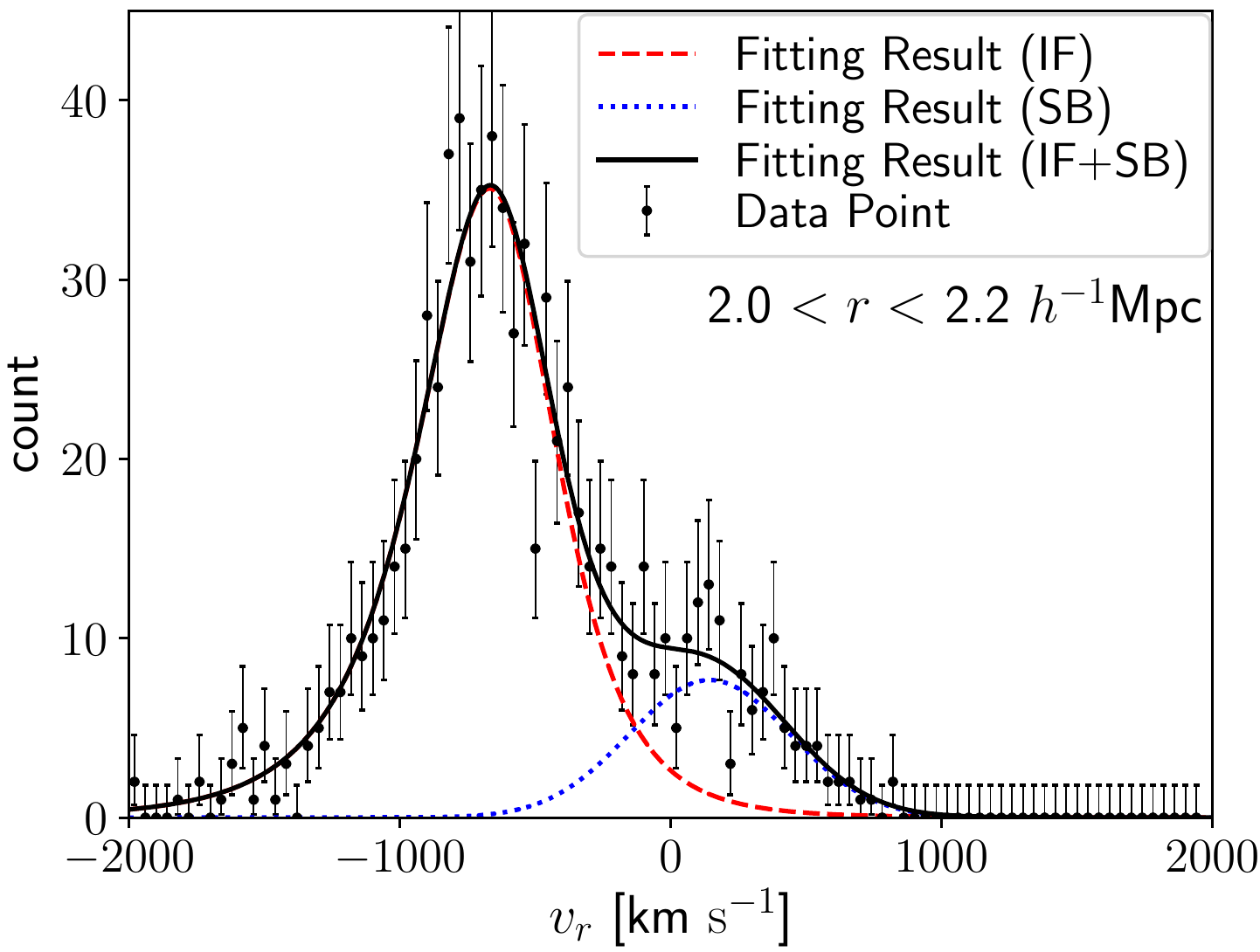}
           \end{minipage}\\
    \begin{minipage}{0.45\textwidth}
    \includegraphics[width=1.0\textwidth]{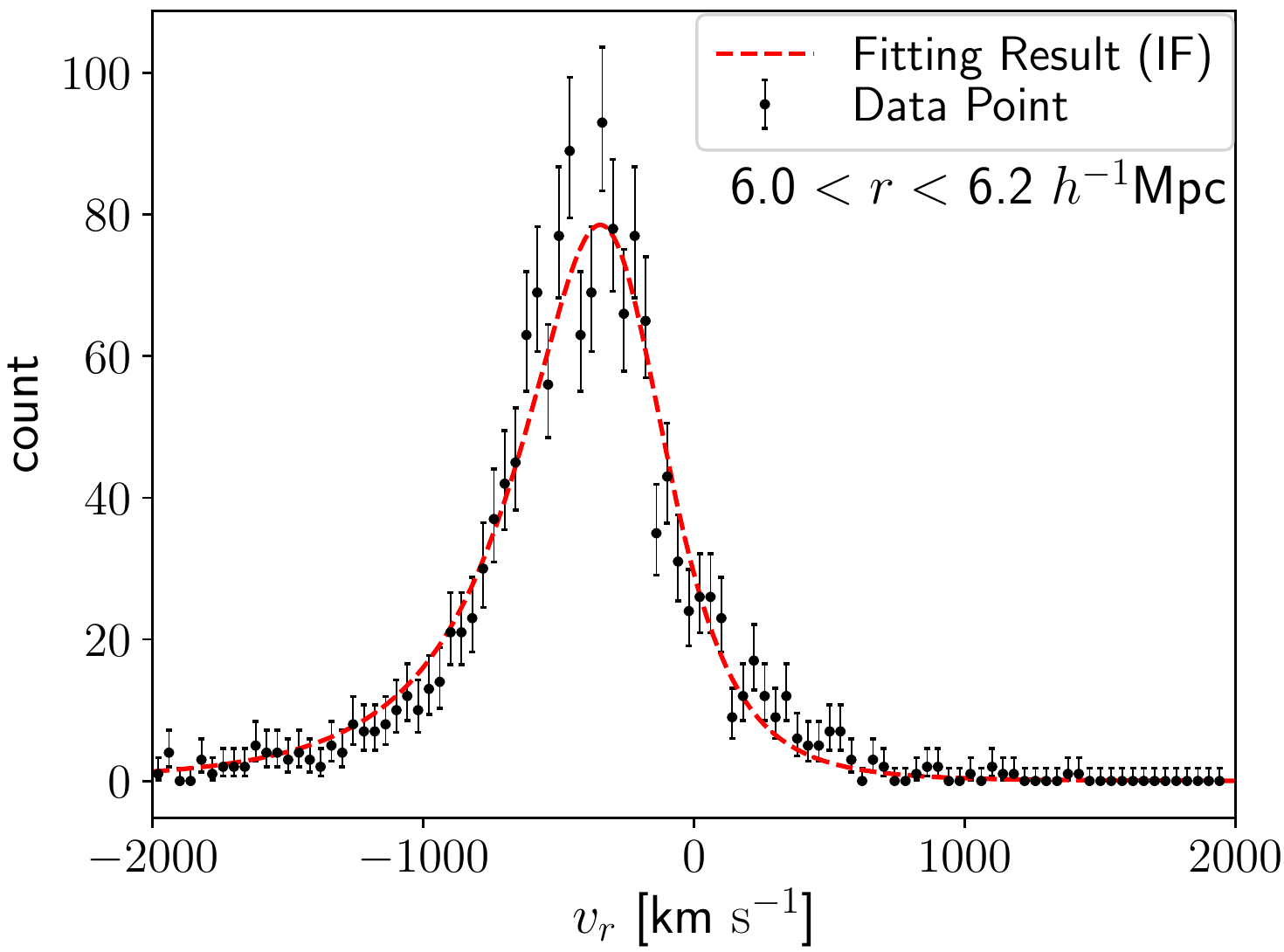}
       \end{minipage}
       \caption{The radial velocity distributions at $2  ~ h^{-1} {\rm{Mpc}}  < r < 2.2  ~ h^{-1} {\rm{Mpc}} $ ($\it{top \ pannel}$) and $6  ~ h^{-1} {\rm{Mpc}} < r < 6.2  ~ h^{-1} {\rm{Mpc}} $ ($ \it{bottom \ pannel}$). The mass range of clusters is $2.125\times 10^{14} ~ h^{-1} M_{\odot} < M_{\rm{cl}} < 2.298 \times 10^{14} ~ h^{-1} M_{\odot} $. Points with error bars are the histogram of radial velocities from our $N$-body simulations, the red dashed line is the best fit line of the infall component, the blue dotted line is the best fit line of the splashback component, and the black solid line is the sum of the dashed and dotted lines.
In the bottom panel, we show only the infall component because there is no splashback component at such large $r$.}

    \label{Fig2}
\end{figure}

Fig. \ref{Fig2} shows examples of radial velocity distributions.
Here the error bars are Poisson errors from the number of haloes in each $v_r$ bin.
The Figure indicates that our model of the radial velocity distribution is in good agreement with the histogram from our $N$-body simulations.
Since the splashback component is negligibly small at large $r$, in fitting we always fix $\alpha = 0$ at $r$ larger than $5  ~ h^{-1} {\rm{Mpc}}$.

\subsubsection{Tangential Velocity Distribution}
Next we model the tangential velocities.
We adopt the Johnson's SU-distribution \citep{Jhonson1949} again as a model function for the tangential velocity distribution of the infall component, and the Gaussian distribution for the splashback component.
Since we assume the spherically symmetric phase space distribution, the tangential velocity distribution must be symmetric aruond $v_{t} = 0$.
Hence, for the tangential velocity distribution, we fix $\gamma_{t} = 0$, and $\xi_{t} = 0$ in the Johnson's SU-distribution and $ \mu_{t} = 0$ for the Gaussian distribution to ensure the symmetry
 \begin{equation}
 \label{eq2-8}
  \begin{split}
p_{v_{t},\rm{infall}} & (v_{t} , r) =  SU (v_{t};\delta_{t},\lambda_{t} ) \\
 =\ & \frac{\delta_{t} }{\lambda_{t} \sqrt{2 \pi} \sqrt{ \{ z(v_{t}) \}^2 + 1}} \exp{ \left[ - \frac{\delta_{t}}{2}  \sinh^{-1}{z(v_{t})}  \right]},
 \end{split}
\end{equation}
where
 \begin{equation}
 \label{eq2-9}
z(v_{t}) = \frac{v_{t} }{ \lambda_{t}} \ ,
\end{equation}
\begin{equation}
 \label{eq2-10}
p_{v_{t},\rm{SB}} (v_{t} , r) = G (v_{t} ;  \sigma_{t}^{2} ) = \frac{1}{\sqrt{2 \pi \sigma_{t}^2 }} \exp{ \left( - \frac{ v_{t}^ 2}{ 2 \sigma_{t}^2 } \right) } \ .
\end{equation}
Again, we note that $\delta_{t}$, $\lambda_{t}$, and $\sigma_{t}$ are functions of $r$.

The model function for the tangential velocity distribution is
 \begin{equation}
   \begin{split}
 \label{eq2-11}
p_{v_{t}} (v_{t} , r) = & \ \int d v_{r} (1 - \alpha )  p_{v_{r},\rm{infall}} (v_{r} ,r)  p_{v_{t},\rm{infall}} (v_{t} ,r) \\
& + \alpha   p_{v_{r},\rm{SB}} (v_{r}  ,r)  p_{v_{t},\rm{SB}} (v_{t}  ,r) \\
= & \ (1 - \alpha)  p_{v_{t},\rm{infall}} + \alpha p_{v_{t},\rm{SB}}\ .
  \end{split}
\end{equation}
In equation (\ref{eq2-11}), $\alpha$ is always fixed to the value obtained from the analysis of the radial velocity distribution, otherwise it is usually very difficult to separate the two components from the tangential velocities alone.
Hence, equation (\ref{eq2-11}) contains additional three free parameters.

\begin{figure}
 \begin{minipage}{0.45\textwidth}
    \includegraphics[width=1.0\textwidth]{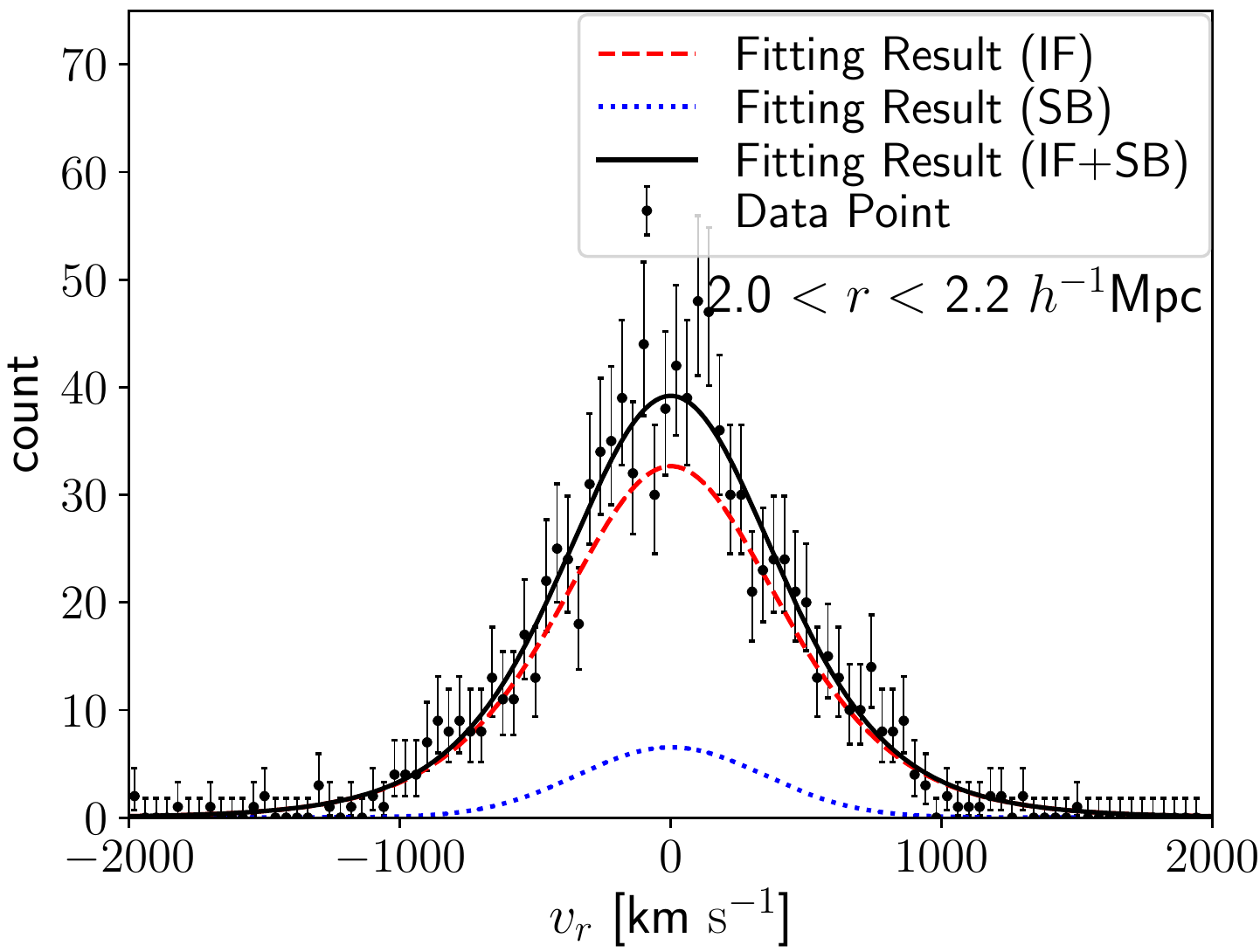}
           \end{minipage}\\
    \begin{minipage}{0.45\textwidth}
    \includegraphics[width=1.0\textwidth]{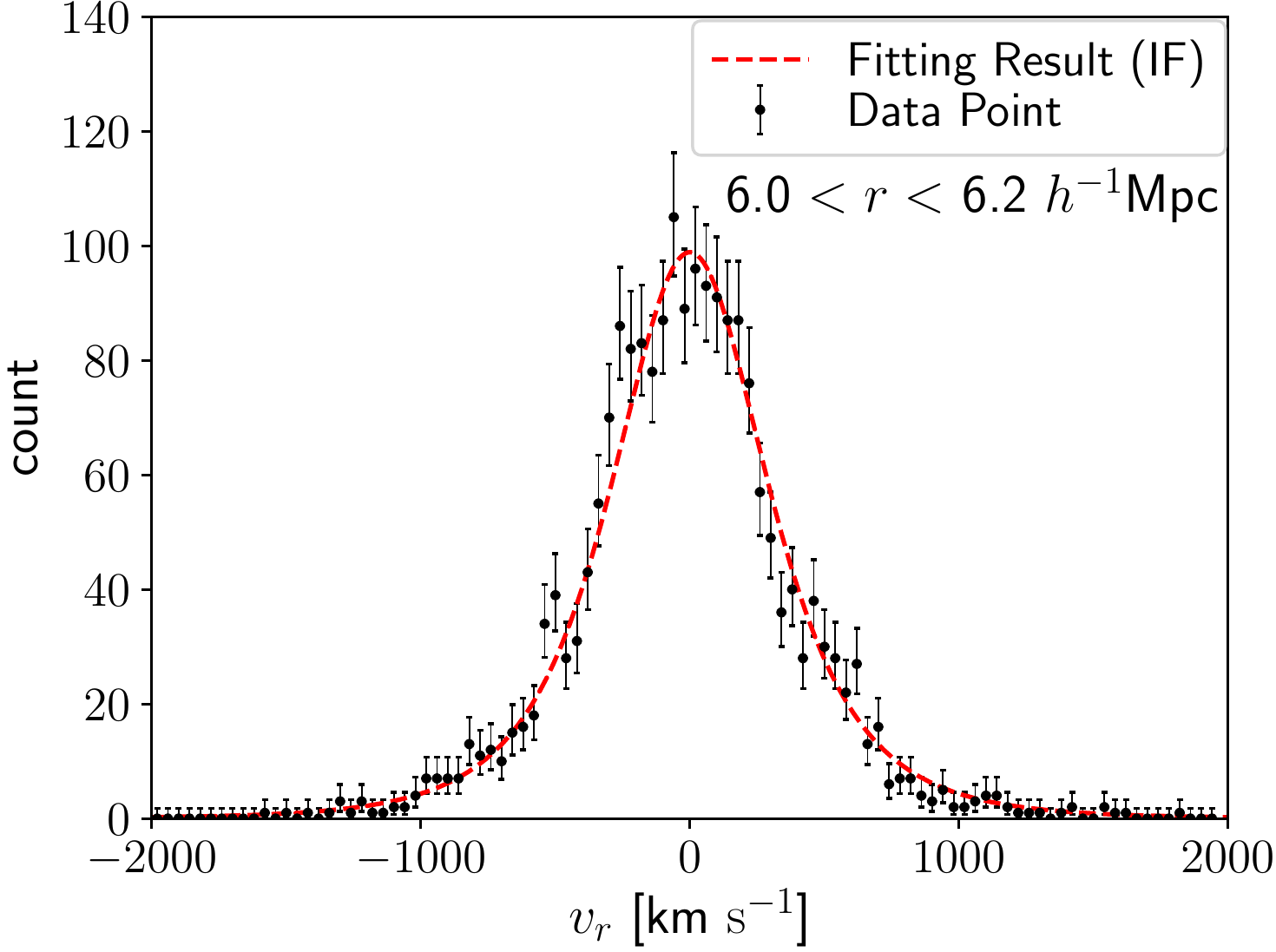}
       \end{minipage}
       \caption{Same as Fig. \ref{Fig2}, but for the tangential velocity distributions.}

    \label{Fig3}
\end{figure}

Fig. \ref{Fig3} shows examples of the tangential velocity distributions.
We can see that our model function for the tangential velocity distribution is also in good agreement with the histogram from our $N$-body simulations.

\subsection{Radial and Cluster Mass Dependences}
\label{S2.3}
By fitting histograms of radial and tangential velocity distributions from $N$-body simulations with equations (\ref{eq2-7}) and (\ref{eq2-11}), we can obtain 10 parameters in these equations for a given $r$ and $M_{\rm cl}$.
In order to construct a model of the three-dimensional phase space distribution as a smooth function of $r$ and $M_{\rm cl}$, we fit these ten parameters with analytic functions. 
Because the four parameters of Johnson's SU-distribution are highly degenerate, we impose some additional assumptions.

First, we assume that $\delta_{r} $, $\delta_{t}$, and $\gamma_{r}$ depend only on $r$ and have no $M_{\rm{cl}}$ dependence. 
We fit $\delta_{r} $, $\delta_{t}$, and $\gamma_{r}$ with analytic functions of $r$ chosen to match the shape obtained from the fitting in Section \ref{S2.2}.
For $\delta_{r} $, $\delta_{t}$, and $\gamma_{r}$, we choose the model functions as 
 \begin{equation}
 \label{eq2-12}
 \begin{split}
\delta_{r, t} (r) = & A_{\delta_{r , t} , 1} \exp{( -  A_{\delta_{r, t} , 2} r )} + A_{\delta_{r , t} , 3} \frac{ r}{ (r + A_{\delta_{r, t} , 4})^{A_{\delta_{r , t} , 5}} } \\
& + A_{\delta_{r , t} , 6} \ ,
\end{split}
\end{equation}
 \begin{equation}
 \label{eq2-13}
 \begin{split}
\gamma_{r } (r) = & A_{\gamma_{r } , 1}  r^{ A_{\gamma_{r } , 2}} + A_{\gamma_{r } , 3} \exp{(  A_{\gamma_{r } , 4} r )} \\
&+ A_{\gamma_{r } , 5} r + A_{\gamma_{r },6} \ ,
\end{split}
\end{equation}
where $A_{\delta_{r, t} , i}$ and $A_{\gamma_{r} , i}$ are free parameters.
After calculating $\delta_{r} $, $\delta_{t}$, and $\gamma_{r}$, we refit the histograms from $N$-body simulations with equations (\ref{eq2-7}) and (\ref{eq2-11}) by replacing these parameters to the smooth functions derived above.
Next, we fit the remaining parameters with analytic functions of $r$ for each galaxy cluster mass bin. The model functions are
\begin{equation}
\label{eq2-14}
\alpha (r) = \frac{-1}{4} \left[ \tanh{ \left( \frac{r -   A_{\alpha ,2}}{ A_{\alpha ,1}}   \right) } - 1 \right] \ ,
\end{equation}
\begin{equation}
\label{eq2-15}
 x  =  ( A_{  x, 1 } - A_{  x, 4 }) r^{ A_{ x , 2} }  + A_{x , 3} r + A_{ x, 4 } \ ,
\end{equation}
\begin{equation}
\label{eq2-16}
 y =    A_{ y, 1 } \exp{ \left( - A_{ y, 2 } r \right ) } +  A_{y, 3 }( r - 2) +  A_{ y, 4 } \ ,
\end{equation}
\begin{equation}
\label{eq2-16-2}
 \sigma_{t} =    A_{ \sigma_{t}, 1 } \left \{ \exp{ \left( - A_{ \sigma_{t}, 2 } r \right )  - 1} \right \}+  A_{\sigma_{t}, 3 } r  +  A_{ \sigma_{t}, 4 } \ ,
\end{equation}
where $A_{ \alpha , i}$, $A_{x, i}$, and $A_{ \sigma_{t}, i}$ are free parameters, and $x$ runs over $\xi_{r}$ and $\mu_{r}$, and $y$ runs over $\lambda_{r}$, $\sigma_{r}$, and $\lambda_{t}$.
We cannot determine parameters of the splashback component where there is no splashback component, i.e. $\alpha \ll 1$.
Hence, for $\alpha$, $\mu_{r}$, $\sigma_{r}$, and $\sigma_{t} $, we need to set the upper limit of the fitting range.
We set the upper limit as the minimum radius where $\alpha$ derived by fitting radial velocity distributions with equation (\ref{eq2-14}) become smaller than 0.1.
We also set the lower limit of the fitting range for computational reasons.
We set the lower limit as the half of the upper limit radius.
Finally, we fit these parameters with one common functional form for the mass dependence
\begin{equation}
\label{eq2-17}
A_{ z , i} =  B_{ z,i , 1} {(M_{{\rm{cl}}} / M_{0})}^{B_{ z,i , 2}} + B_{ z ,i, 3} \ ,
\end{equation}
where $M_{0}$ is a pivot mass,  $B_{ z , i ,l}$ are free parameters, and $z$ runs over $\alpha$, $\xi_{r}$, $\lambda_{r}$, $\mu_{r}$, $\sigma_{r}$, $\lambda_{t}$, and $\sigma_{t}$.
We adopt the pivot mass $M_{0} = 1.429 \times 10^{13} h^{-1} M_{\odot} $.

\begin{figure}
 \begin{minipage}{0.45\textwidth}
    \includegraphics[width=1.0\textwidth]{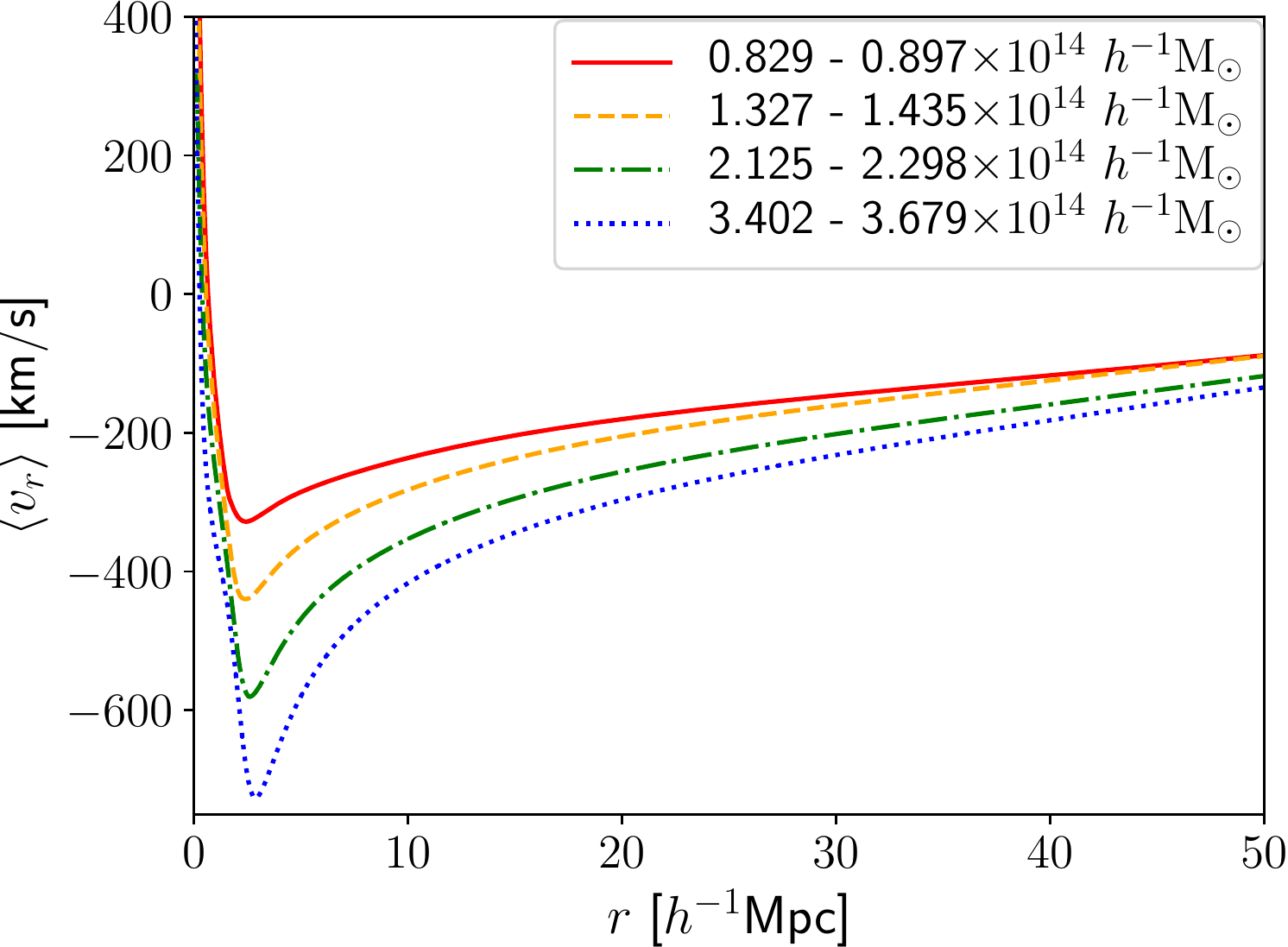}
           \end{minipage}\\
    \begin{minipage}{0.45\textwidth}
    \includegraphics[width=1.0\textwidth]{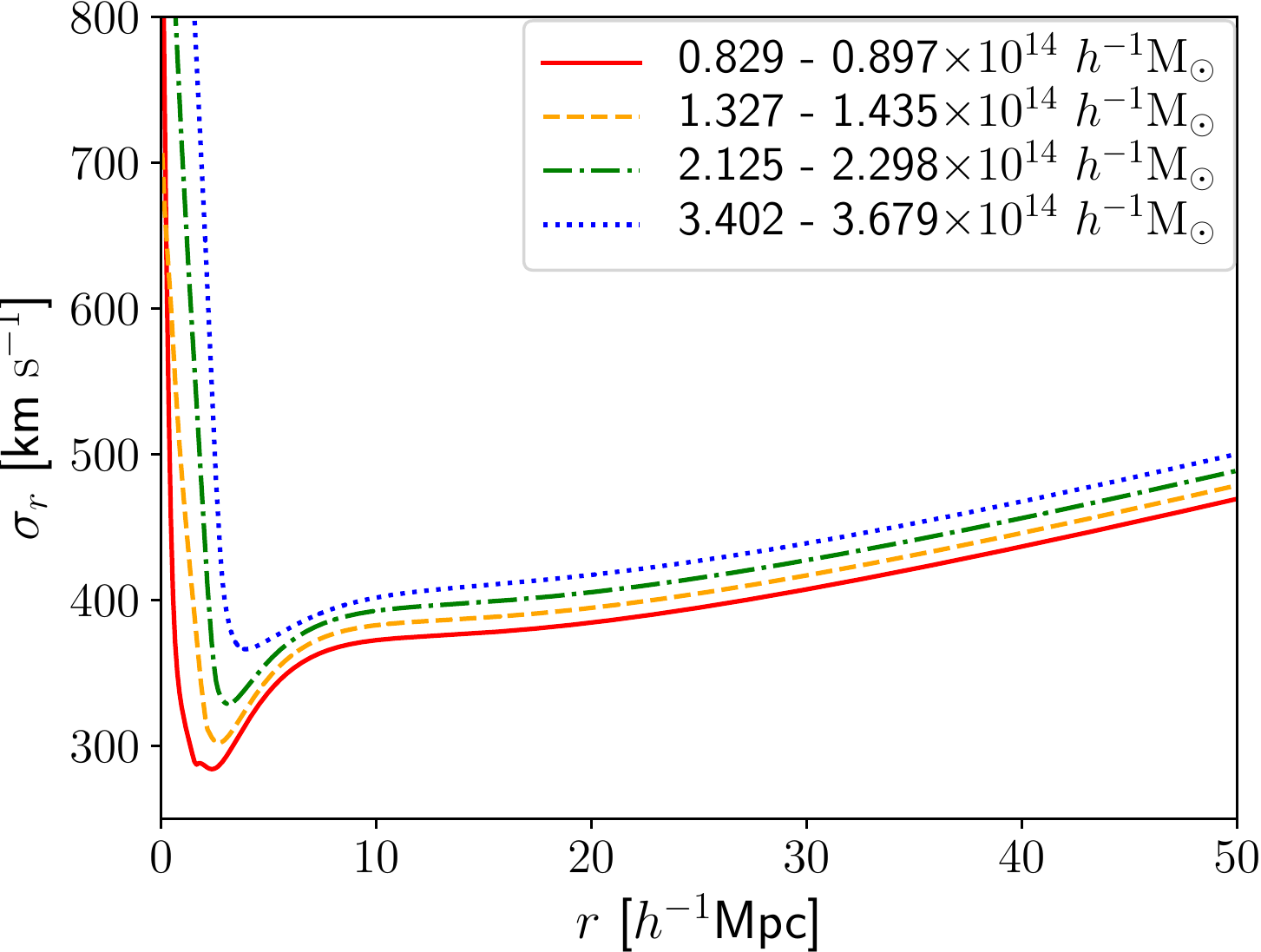}
       \end{minipage}\\
     \begin{minipage}{0.45\textwidth}
   \includegraphics[width=1.0\textwidth]{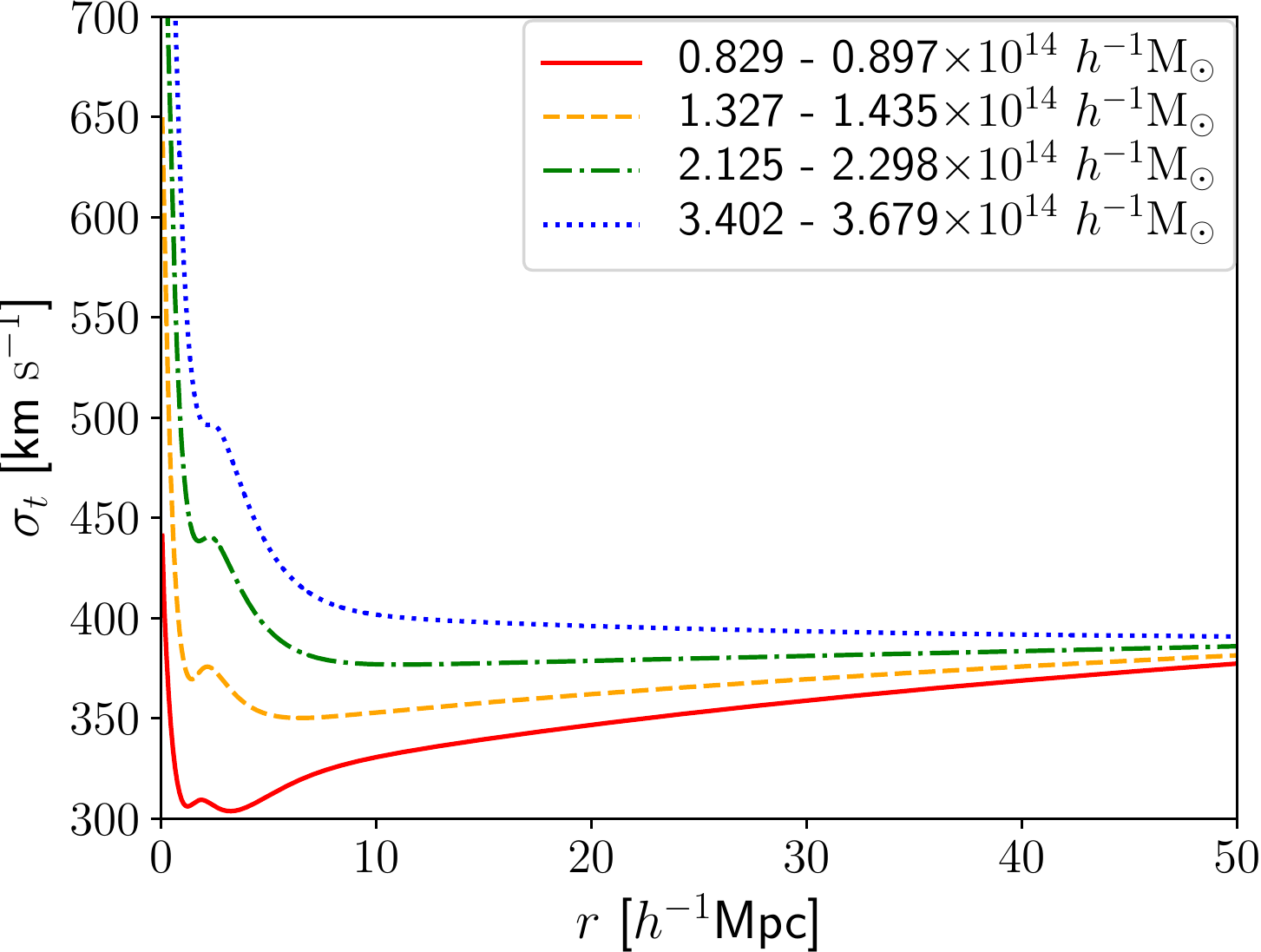}
       \end{minipage}
       \caption{Radial dependences of the average of $v_{r}$, $\langle v_{r} \rangle$ ($ \it{top}$), the standard deviations of $v_{r}$, $\sigma_{r}$ ($ \it{middle}$) and $v_{t}$, $\sigma_{t}$ ($ \it{bottom}$) for four galaxy cluster mass bins $M_{\rm{cl}}$. The different lines show different $M_{\rm{cl}}$.}
    \label{Fig4}
\end{figure}

Now, we can calculate parameters introduced in Section \ref{S2.2} as a smooth functions of $r$ and $M_{\rm{cl}}$.
Hence, by using these parameters, we can compute the dependence of the average of $v_{r}$, $\langle v_{r} \rangle$, the standard deviations of $v_{r}$, $\sigma_{r}$ and $v_{t}$, $\sigma_{t}$ on $r$ and $M_{\rm{cl}}$ as 
\begin{equation}
\label{eq2-18}
\langle v_{r} \rangle = (1 - \alpha) \Bigl[ \xi_{r} - \lambda_{r} \{ \exp{( {\delta_{r}}^{-2} )} \}^{1/2} \sinh{ \frac{\gamma_{r}}{\delta_{r}}} \Bigr] + \alpha \mu_{r} \ ,
\end{equation}
\begin{equation}
\label{eq2-19}
\begin{split}
\sigma_{r}^2 =\  &  (1 - \alpha) \biggl[  \frac{\lambda_{r}^2}{2} \Bigl\{ 2 \exp{( \delta_{r}^{-4} )} \sinh^{2}{ \Bigl( \frac{\gamma_{r}}{\delta_{r}} \Bigr)} +   \exp{( \delta_{r}^{-2} )}  -1 \Bigr\} \\
& -2 \xi_{r} \lambda_{r}  \exp{( \delta_{r}^{-1} )} \sinh{\Bigl( \frac{\gamma_{r}}{\delta_{r}} \Bigr)} + \xi_{r}^{2}  \biggr] \\
& + \alpha (\sigma_{r}^2 + \mu_{r}^2 ) + { \langle v_{r} \rangle }^{2}  \ .
\end{split}
\end{equation}

\begin{equation}
\label{eq2-20}
\sigma_{t}^2 =  (1 - \alpha) \frac{\lambda_{t}^{2}}{2} \{  \exp{(- \delta_{t}^{4})} - 1  \} + \alpha \sigma_{t}^{2}  \ .
\end{equation}
Fig. \ref{Fig4} show the dependence of $\langle v_{r} \rangle$, $\sigma_{r}$, and $\sigma_{t}$ on $r$ and $M_{\rm{cl}}$.
We can see that more massive clusters have lower $\langle v_{r} \rangle$ and higher $\sigma_{r}$, and $\sigma_{t}$, even at $r$ larger than $10 ~ h^{-1} {\rm{Mpc}} $.

\begin{figure}
    \includegraphics[width=\columnwidth]{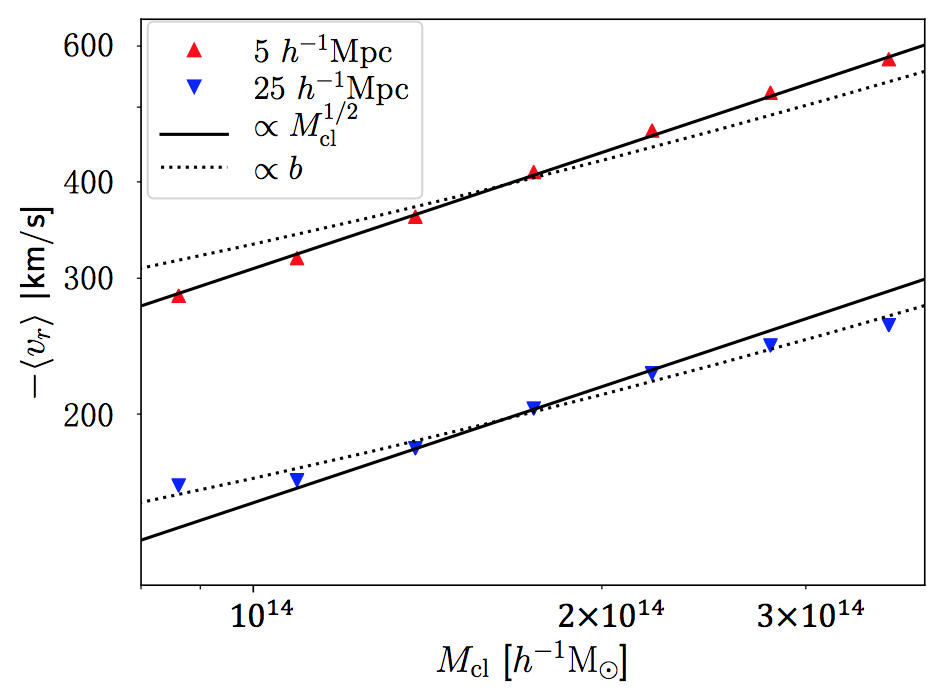}
    \caption{Cluster mass dependence of the average of $v_{r}$, $\langle v_{r} \rangle$ at $ r = 5.0  ~ h^{-1} {\rm{Mpc}} $ ($\it{red ~ triangle ~ up}$) and $r = 25.0  ~ h^{-1} {\rm{Mpc}} $ ($\it{blue ~ triangle ~ down}$).
Solid lines are the best fit lines assuming the mass dependence from equation (\ref{eq2-21}), and the dashed lines are the best fit lines assuming the mass dependence from equation (\ref{eq2-23}), where we adopt a halo bias model of \citet{Bhattacharya2011} to compute $b(M_{\rm cl})$.}
    \label{Fig4-2}
\end{figure}

As we will discuss later, one of the key parameters that constrain cluster masses is the radial velocity $v_r$, whose mass dependence can be understood by a simple analytic argument.
First, when haloes are sufficiently close to clusters, based on the energy conservation, we expect that the radial (i.e., infall) velocity scales as
\begin{equation}
\label{eq2-21}
v_{r} \propto M_{\rm{cl}}^{1/2} \ .
\end{equation}
Next, as shown in \citet{Fisher95} and \citet{Agrawal2017}, at sufficiently large $r$ the radial velocity is derived by using the linear perturbation theory as
\begin{equation}
\label{eq2-22}
v_{r}(r) = - \frac{H f b }{\pi^{2}} \int dk k P_{\rm{mm}}(k) j_{1}{(kr)} \ ,
\end{equation}
where $f = \rm{d} \ln{D} / \rm{d} \ln{a}$ is the logarithmic growth rate with $D$ being the linear growth factor, $b$ is the linear bias factor (e.g., \citealt{Bhattacharya2011}), $P_{\rm{mm}}(k)$ is the matter power spectrum, and $j_{1}{(kr)}$ is the spherical Bessel function of the first order.
This linear perturbation theory result suggests that the radial velocity should scale as
\begin{equation}
\label{eq2-23}
v_{r} \propto b(M_{\rm{cl}}) \ .
\end{equation}
Fig. \ref{Fig4-2} shows the cluster mass dependence of the average of $v_{r}$, $\langle v_{r} \rangle$.
We determine the normalization of equations (\ref{eq2-21}) and (\ref{eq2-23}) so as to fit the points.
We can see that at a small radius the cluster mass dependence of $\langle v_{r} \rangle$ follows equation (\ref{eq2-21}), whereas at a large radius it follows equation (\ref{eq2-23}).
The success of this simple analytic argument implies that we may be able to construct a fully analytic model of the average infall velocity that smoothly connects equations (\ref{eq2-22}) and (\ref{eq2-23}), which we leave for future work.

\section{Projected Phase Space Distribution}
\label{S3.0}

\subsection{Effect of the Hubble Flow}
\label{S3.1}
\begin{figure}
    \includegraphics[width=\columnwidth]{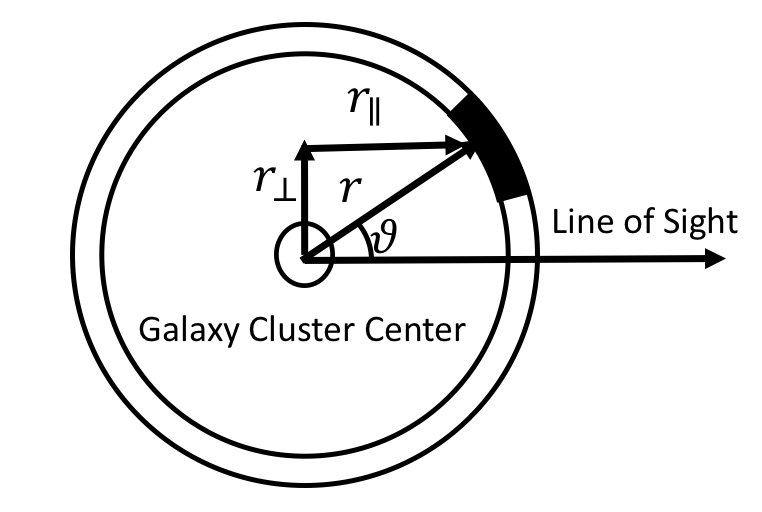}
    \caption{Schematic illustration of the integration parameters in equation (\ref{eq3-1})}
    \label{Fig5}
\end{figure}

Next we derive the PDF of the line-of-sight velocity $v_{\rm los}$ that can directly be compared with observations.
We can do so by projecting the three-dimensional phase space distribution derived in the previous section along the line-of-sight with the effect of the Hubble flow.
In Section \ref{S2.2} and Section \ref{S2.3}, we obtain equation (\ref{eq2-1}) as a smooth function of $r$ and $M_{\rm{cl}}$.
We can describe the PDF of $v_{\rm{los}}$ by using equation (\ref{eq2-1}) as
 \begin{equation}
 \label{eq3-1}
 \begin{split}
& p_{v_{\rm{los}}} (v_{\rm{los}} , r_{\perp})  = \\
 & \frac{1}{N (r_{\perp})} \int^{\infty}_{-\infty} d v_{r}   \int^{\infty}_{-\infty} d v_{t}  \int^{\infty}_{-\infty} d r_{\parallel}   n(r)  p_{v}(v_{r} , v_{t} , r)  \delta_{D} ( v_{\rm{los}} - v'_{\rm{los}} ) ,
\end{split}
\end{equation}
where $r_{\perp}$ is the transverse distance from the cluster centre, $r_{\parallel}$ is the line-of-sight distance from the cluster centre, $n(r)$ is the number density profile of haloes, $\delta_{D} (x)$ is the Dirac's delta function, and $p_{v}(v_{r} , v_{t} , r)$ is the three-dimensional phase space distribution defined in equation (\ref{eq2-1}).
We note that $r$ is defined as 
 \begin{equation}
 \label{eq3-2}
r \equiv \sqrt{r_{\parallel}^2 + r_{\perp}^2} \ ,
\end{equation}
and $N$ is a normalization factor defined as
\begin{equation}
 \label{eq3-3}
 \begin{split}
& N (r_{\perp}) \equiv \\
& \int^{v_{\rm{max}}}_{-v_{\rm{max}}} d v_{\rm{los}} \int^{\infty}_{-\infty} d v_{r}   \int^{\infty}_{-\infty} d v_{t}  \int^{\infty}_{-\infty} d r_{\parallel}  \\
& \int^{\infty}_{-\infty} d v_{\rm{los}}  n(r)  p_{v}(v_{r} , v_{t} , r)  \delta_{D} ( v_{\rm{los}} - v'_{\rm{los}} ) \ ,
\end{split}
\end{equation}
where $v_{\rm{max}}$ is the range of $v_{\rm{los}}$ we calculate.
Because of practical reasons, we set $v_{\rm{max}} = 2000 ~ {\rm{km \ s^{-1}}}$ in this paper.
The choice of $v_{\rm max}$ is unimportant because we can also apply this cut to observational data.
Also, $v'_{\rm{los}}$ is defined as
 \begin{equation}
 \label{eq3-4}
v'_{\rm{los}} \equiv ( \cos \theta ) v_{r} + ( \sin \theta ) v_{t} + \frac{H(z)  r_{\parallel}}{1 + z} ,
\end{equation}
where $H(z)$ is the Hubble parameter, and $\theta$ corresponds to the angle between the line from the cluster centre and the line-of-sight, as shown in Fig. \ref{Fig5}.

In what follows, we set the integration range of $v_{r}$ and $v_{t}$ as $| v_{r} |,| v_{t}| < 2000 ~ \rm{km \ s^{-1}}$, because we find that there is almost no probability out of this range in the three-dimensional phase space distribution at any radii of our interest.
We also set the integration range of $r_{\parallel}$ as $-40 ~ h^{-1} {\rm{Mpc}}  < r_{\parallel}  < 40 ~ h^{-1} {\rm{Mpc}}$, because for our choice of $v_{\rm los}<2000 ~ {\rm km\ s^{-1}}$, this integration range is sufficiently large.

Since we construct the three-dimensional phase space distribution ($p_{v}$) only at $r$ larger than the lower limit we set in Section \ref{S2.3}, we cannot describe the PDF of $v_{\rm{los}}$ at $r_{\perp}$ less than the lower limit. 
Since in this paper we focus on the PDF of $v_{\rm los}$ at radii larger than 2$h^{-1}$ Mpc that is larger than the lower limit, this lower limit does not affect our analysis.

\begin{figure}
 \begin{minipage}{0.45\textwidth}
    \includegraphics[width=1.0\textwidth]{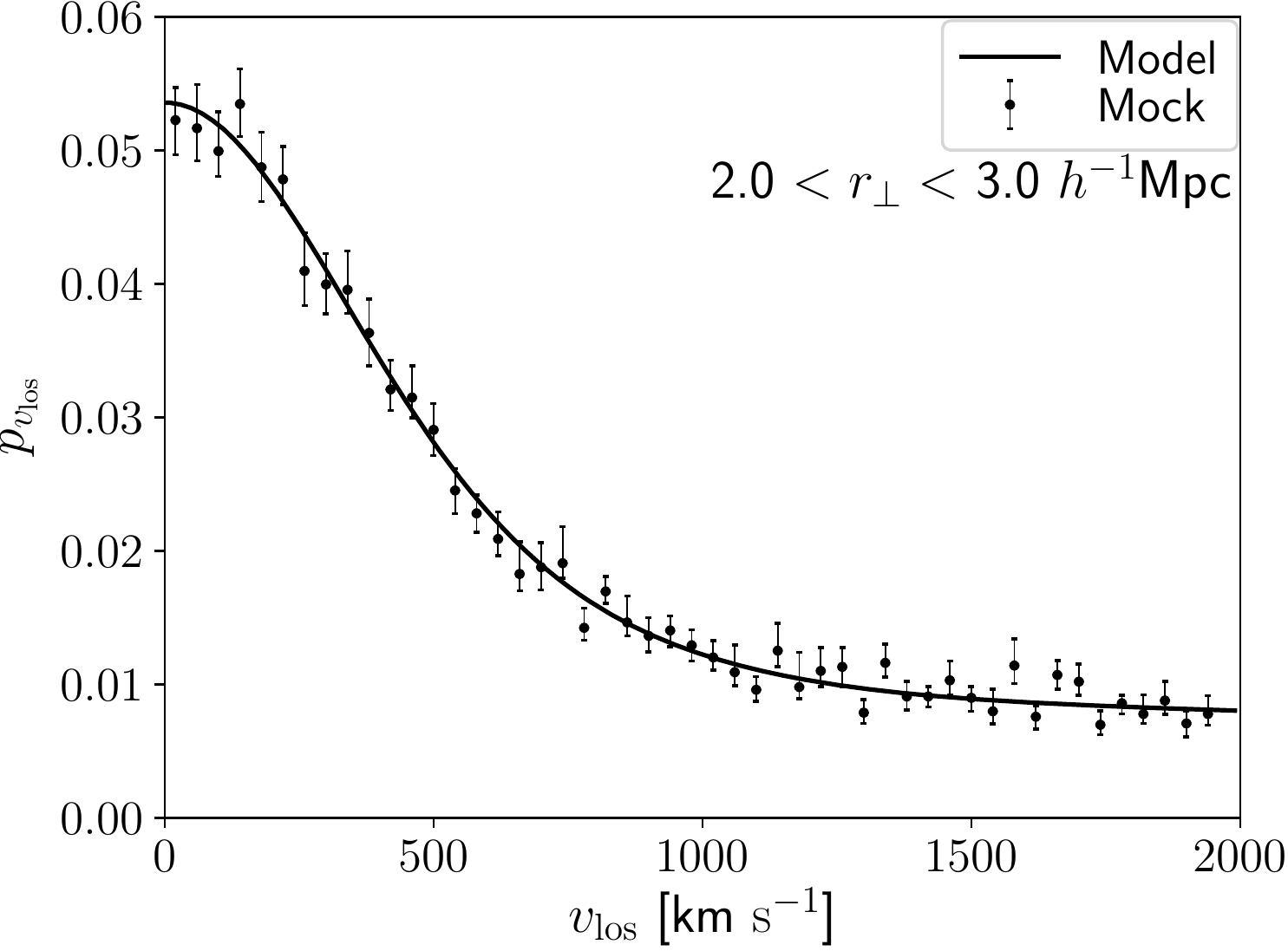}
           \end{minipage}\\
    \begin{minipage}{0.45\textwidth}
    \includegraphics[width=1.0\textwidth]{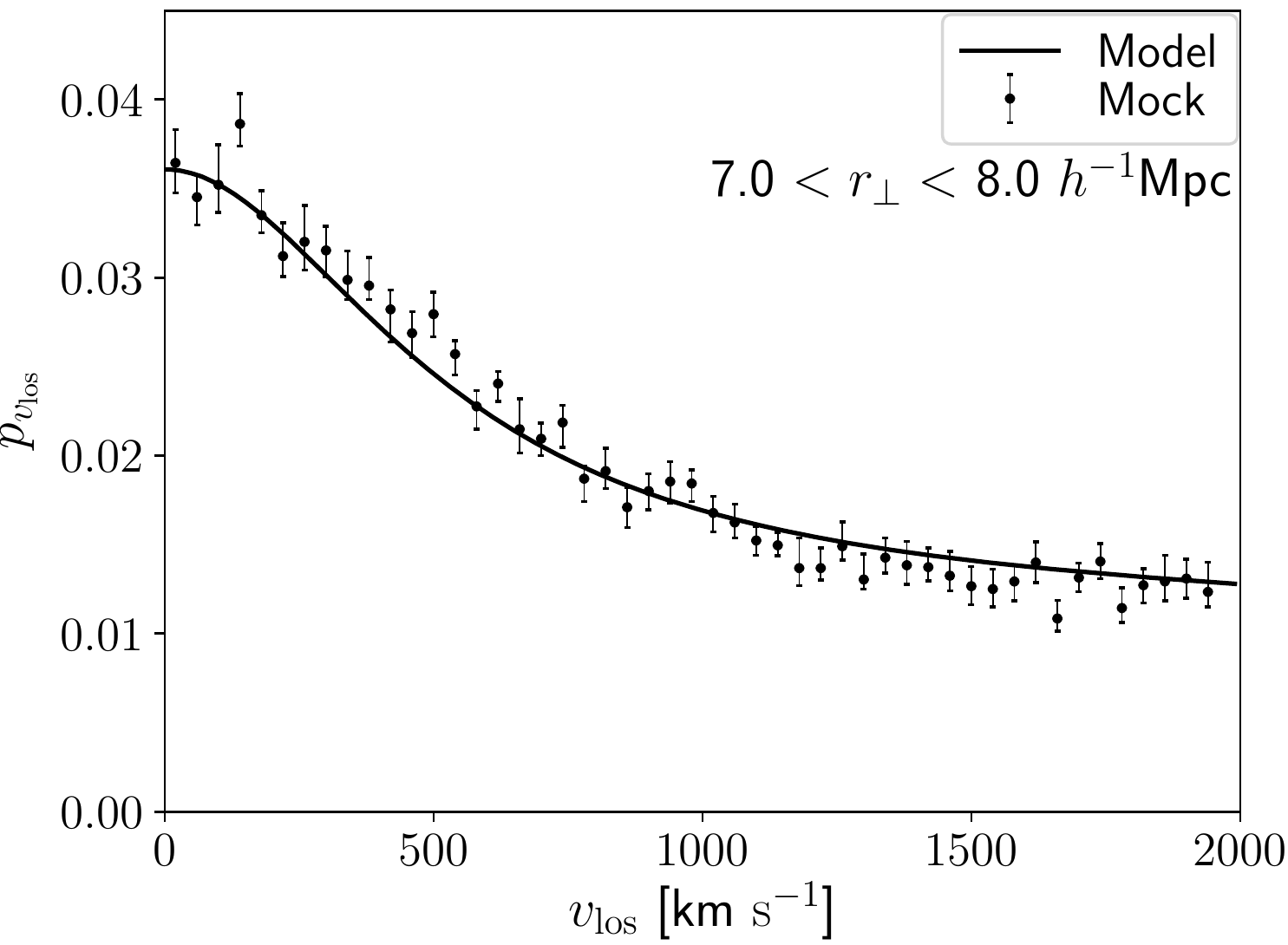}
       \end{minipage}
       \caption{Comparison of PDFs of $v_{\rm{los}}$ calculated from our model and those from mock observation of our $N$-body simulations for $2.125 \times 10^{14} ~ h^{-1} M_{\odot} < M_{{\rm{cl}}} < 2.298 \times 10^{14} ~ h^{-1} M_{\odot} $ at $2  ~  h^{-1} {\rm{Mpc}} < r_{\perp} < 3 ~ h^{-1} {\rm{Mpc}} $ ($\it{top}$) and $7  ~ h^{-1} {\rm{Mpc}} < r_{\perp} < 8 ~ h^{-1} {\rm{Mpc}} $ ($ \it{bottom}$). 
Solid lines are $p_{v_{\rm{los}}}$ calculated from equation (\ref{eq3-1}), and points show PDFs of $v_{\rm{los}}$ from mock observation of our $N$-body simulations.
$\chi^2 / \rm{dof} = 0.97$ for the top panel and $\chi^2 / \rm{dof} = 1.40$ for the bottom panel, where the degree of freedom is 50.
We use 6,419 (18,715) haloes for the top (bottom) panel.}
    \label{Fig6}
\end{figure}

In the analysis of observed PDF of $v_{\rm los}$, one should adopt $n(r)$ that is estimated directly from the observation.
However,  in comparison with the $N$-body simulation results we use $n(r)$ directly measured in the $N$-body simulations.
We now check the accuracy of our model calculation of PDFs of $v_{\rm los}$.
We do so by comparing the PDF of $v_{\rm los}$ computed from equation (\ref{eq3-1}) with the PDF directly measured from our $N$-body simulations.
In order to allow accurate comparisons, here we estimate the error bars of $p_{v_{\rm{los}}}$ directly from our $N$-body simulations, as described as follows.
First, we divide our $N$-body simulations into thirty-two sub-regions.
We then measure $p_{v_{\rm{los}}}$ in each sub-regions and calculate the sample variances of $p_{v_{\rm{los}}}$ for each sub-regions by comparing $p_{v_{\rm{los}}}$ from the variance of $p_{v_{\rm los}}$ values among the thirty two sub-regions.
We find that errors estimated by this method are similar to Poisson errors, but are slightly larger than Poisson errors typically by 16\%.
This small difference of errors may arise from the contribution from the large-scale structure of the Universe.
For simplicity, we also ignore the covariance between different $v_{\rm los}$ bins, as we do not find any significant covariance between different bins in our sub-region analysis.
In Fig. \ref{Fig6}, we compare $p_{v_{\rm{los}}}$ calculated in equation (\ref{eq3-1}) with $p_{v_{\rm{los}}}$ directly measured from our $N$-body simulations.
We obtain relatively good $\chi^2 / \rm{dof}$, which implies that our model is reasonably accurate and the estimated errors from the sub-regions are also reasonable.

\subsection{Cluster Mass Dependence}
\label{S3.2}
\begin{figure*}
 \begin{minipage}{0.31\textwidth}
    \includegraphics[width=1.0\textwidth]{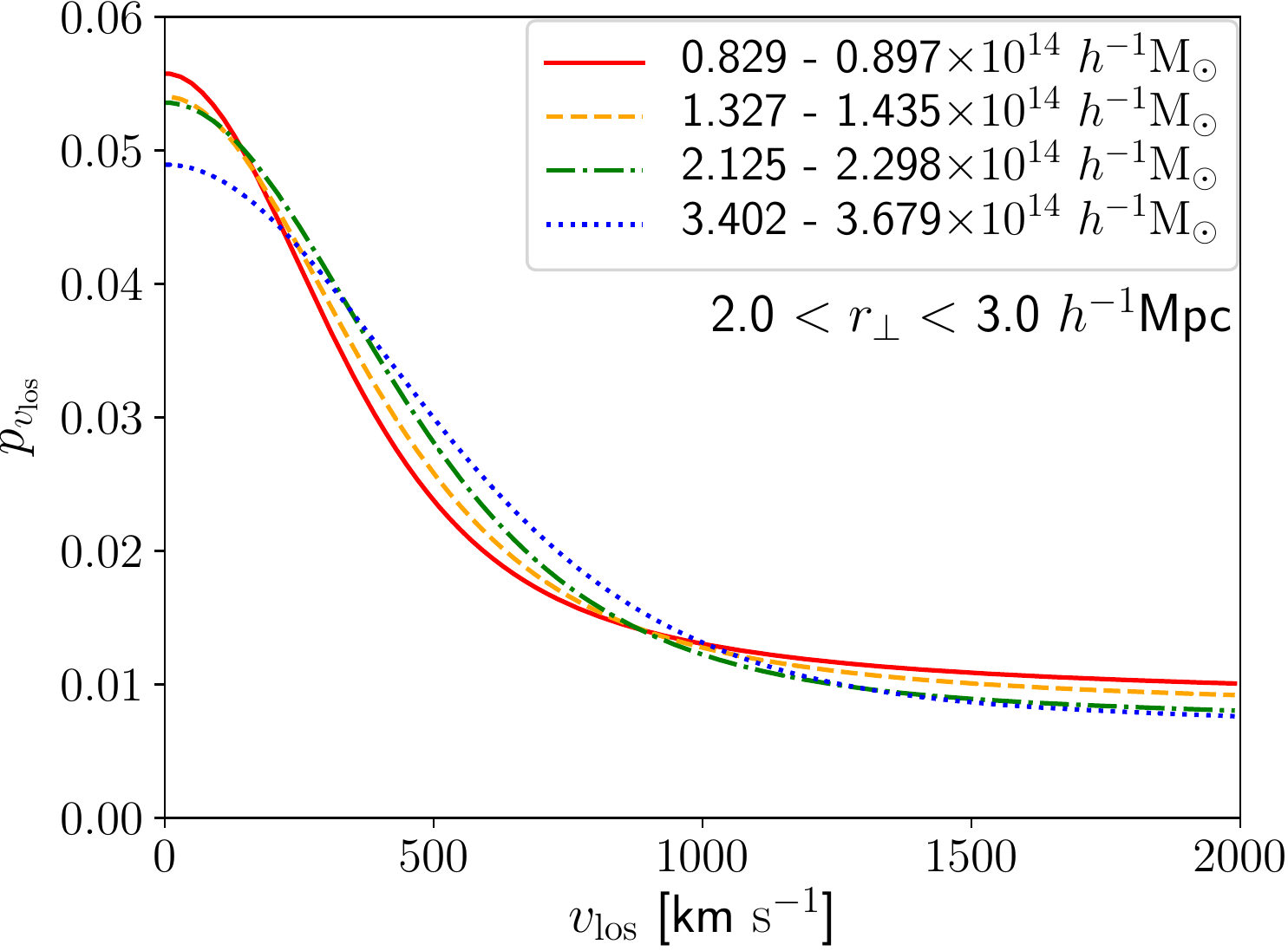}
           \end{minipage}
    \begin{minipage}{0.31\textwidth}
    \includegraphics[width=1.0\textwidth]{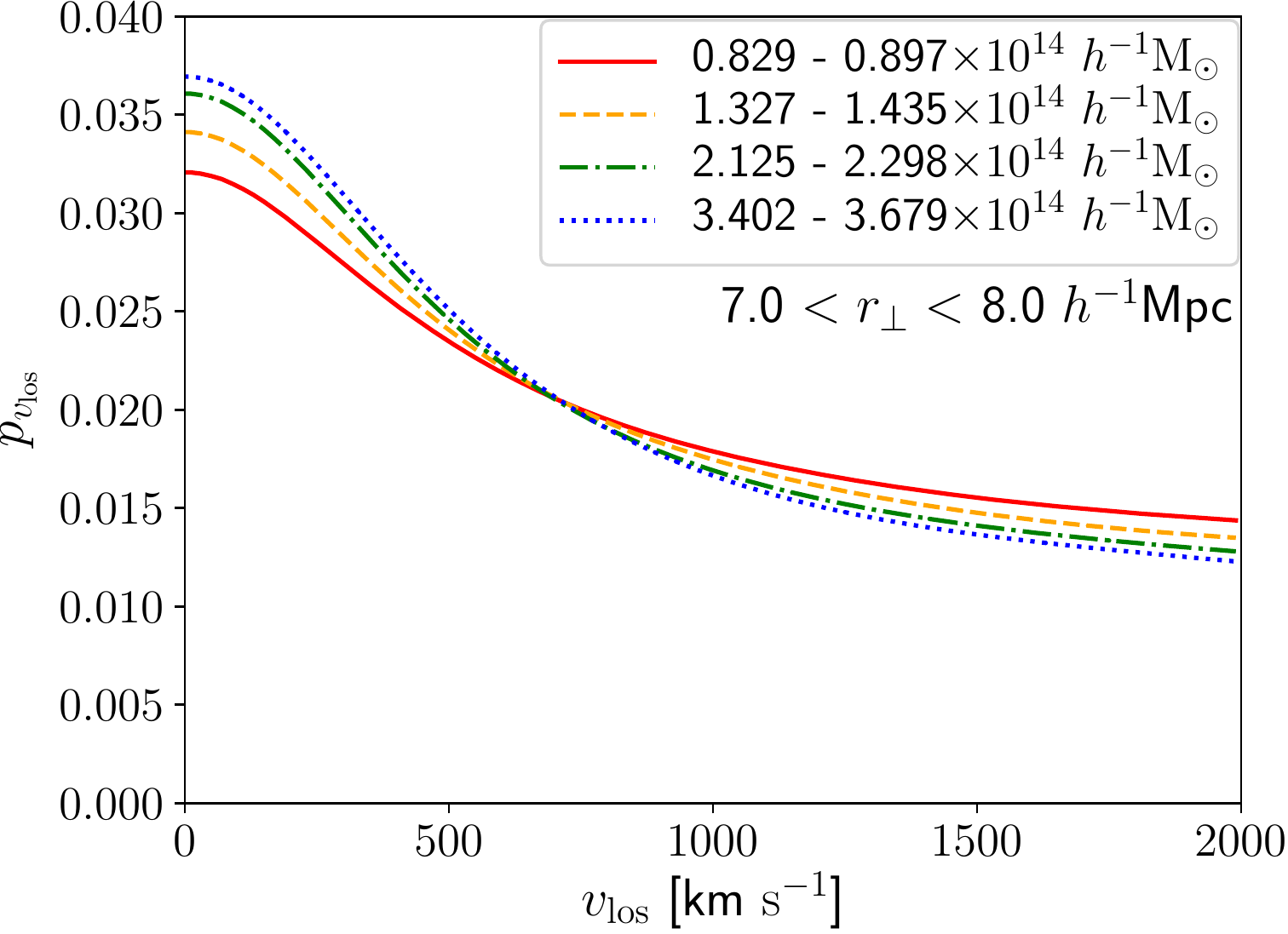}
       \end{minipage}
     \begin{minipage}{0.31\textwidth}
   \includegraphics[width=1.0\textwidth]{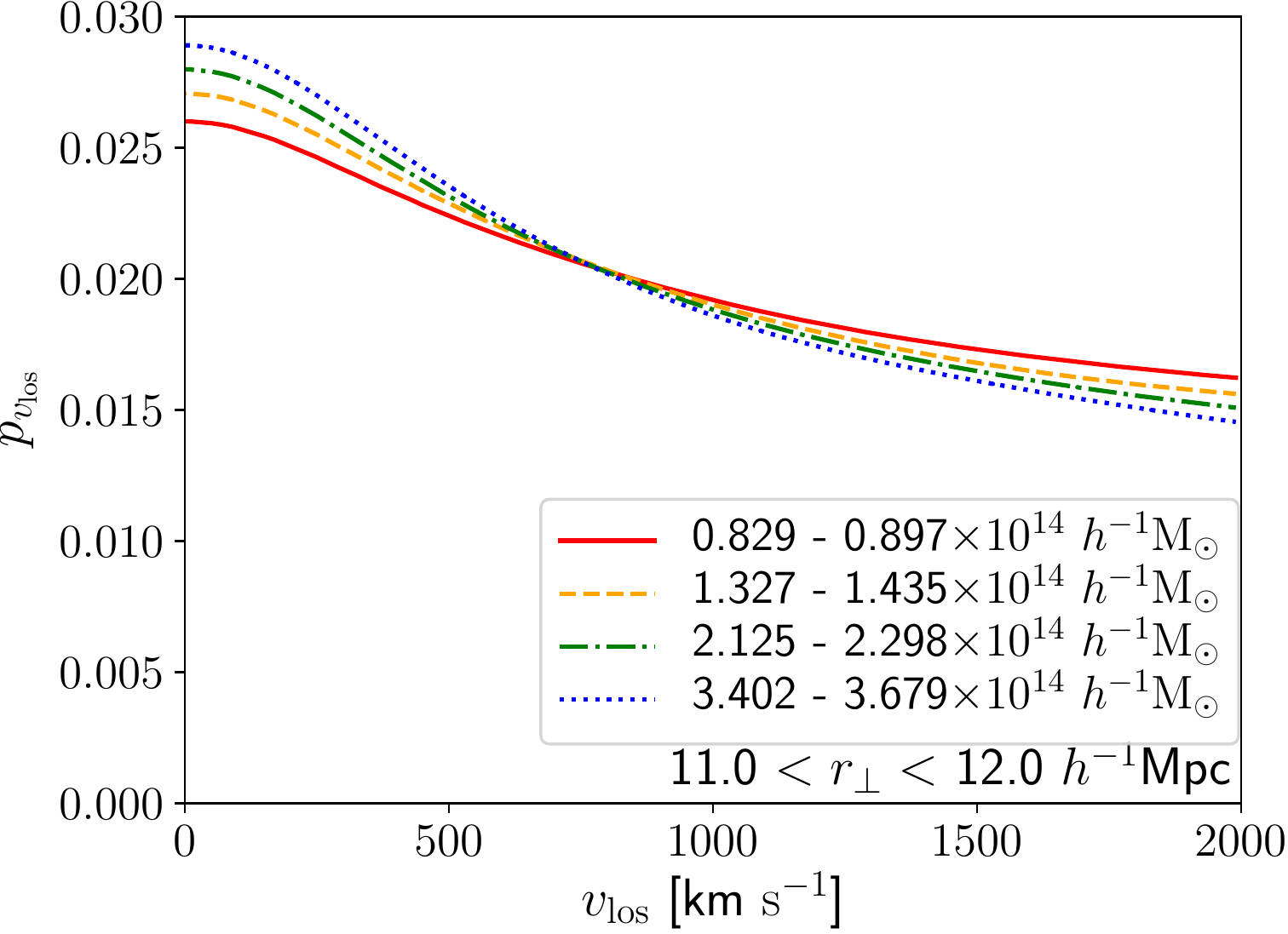}
       \end{minipage}
       \caption{PDF of $v_{\rm{los}}$ calculated from our model at $2  ~ h^{-1} {\rm{Mpc}} < r_{\perp} < 3 ~ h^{-1} {\rm{Mpc}} $ ($\it{left}$), $7  ~ h^{-1} {\rm{Mpc}} < r_{\perp} < 8 ~ h^{-1} {\rm{Mpc}} $ ($ \it{middle}$), and $11  ~ h^{-1} {\rm{Mpc}} < r_{\perp} < 12 ~ h^{-1} {\rm{Mpc}} $ ($ \it{right}$) for six different $M_{\rm{cl}}$. The different lines show PDFs for different $M_{\rm{cl}}$.}
    \label{Fig7}
\end{figure*}

\begin{figure*}
 \begin{minipage}{0.31\textwidth}
    \includegraphics[width=1.0\textwidth]{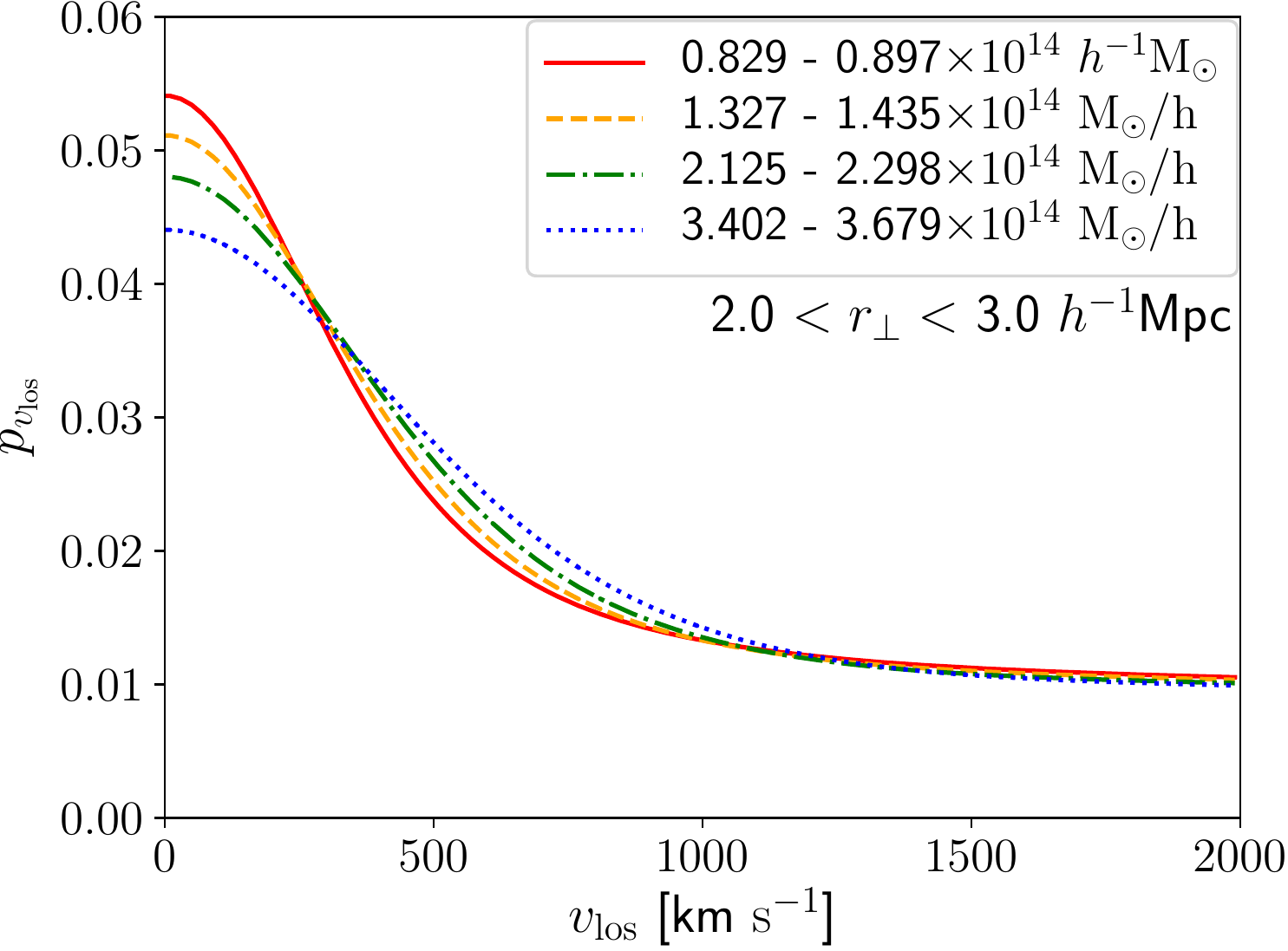}
           \end{minipage}
    \begin{minipage}{0.31\textwidth}
    \includegraphics[width=1.0\textwidth]{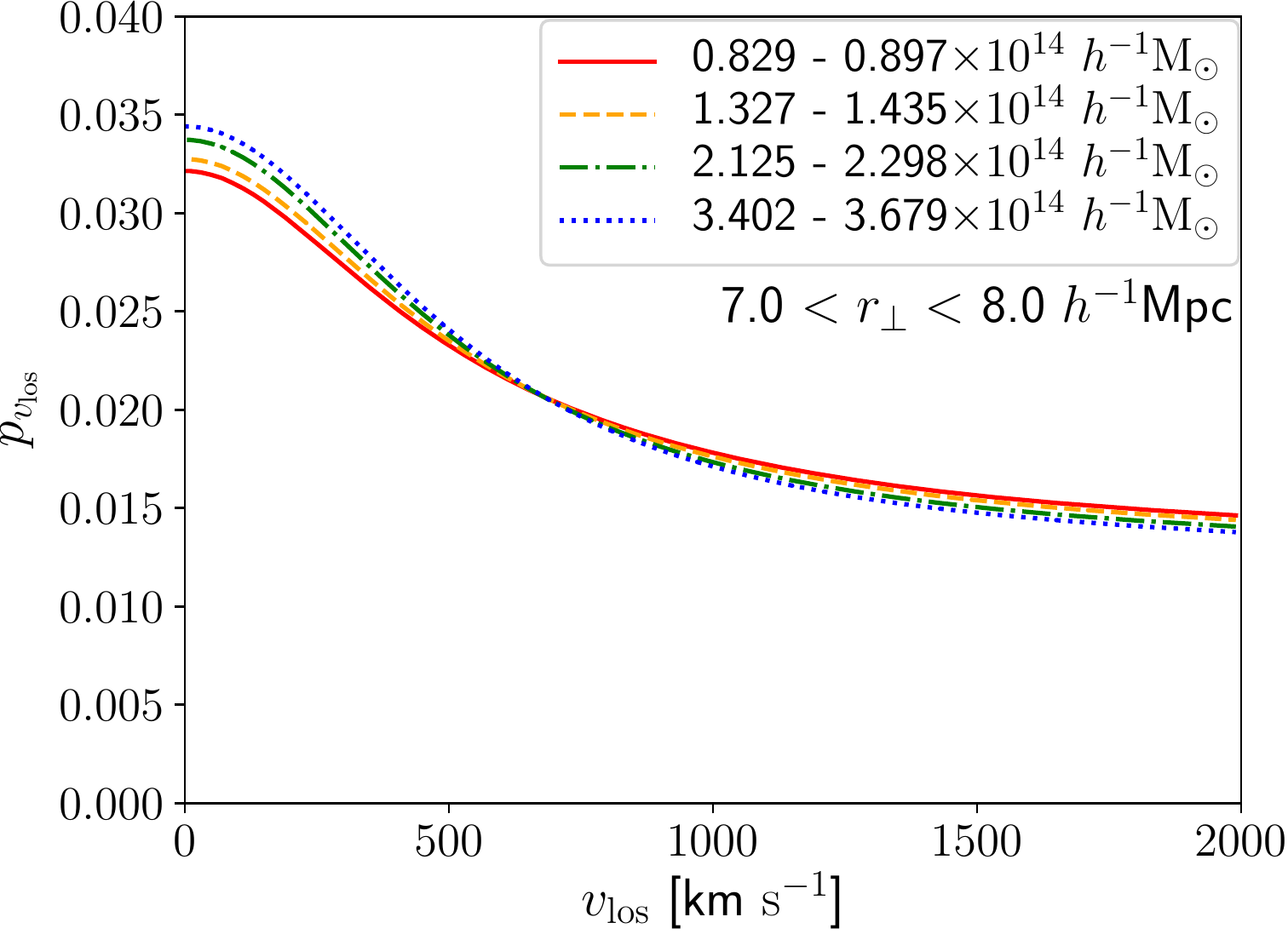}
       \end{minipage}
     \begin{minipage}{0.31\textwidth}
   \includegraphics[width=1.0\textwidth]{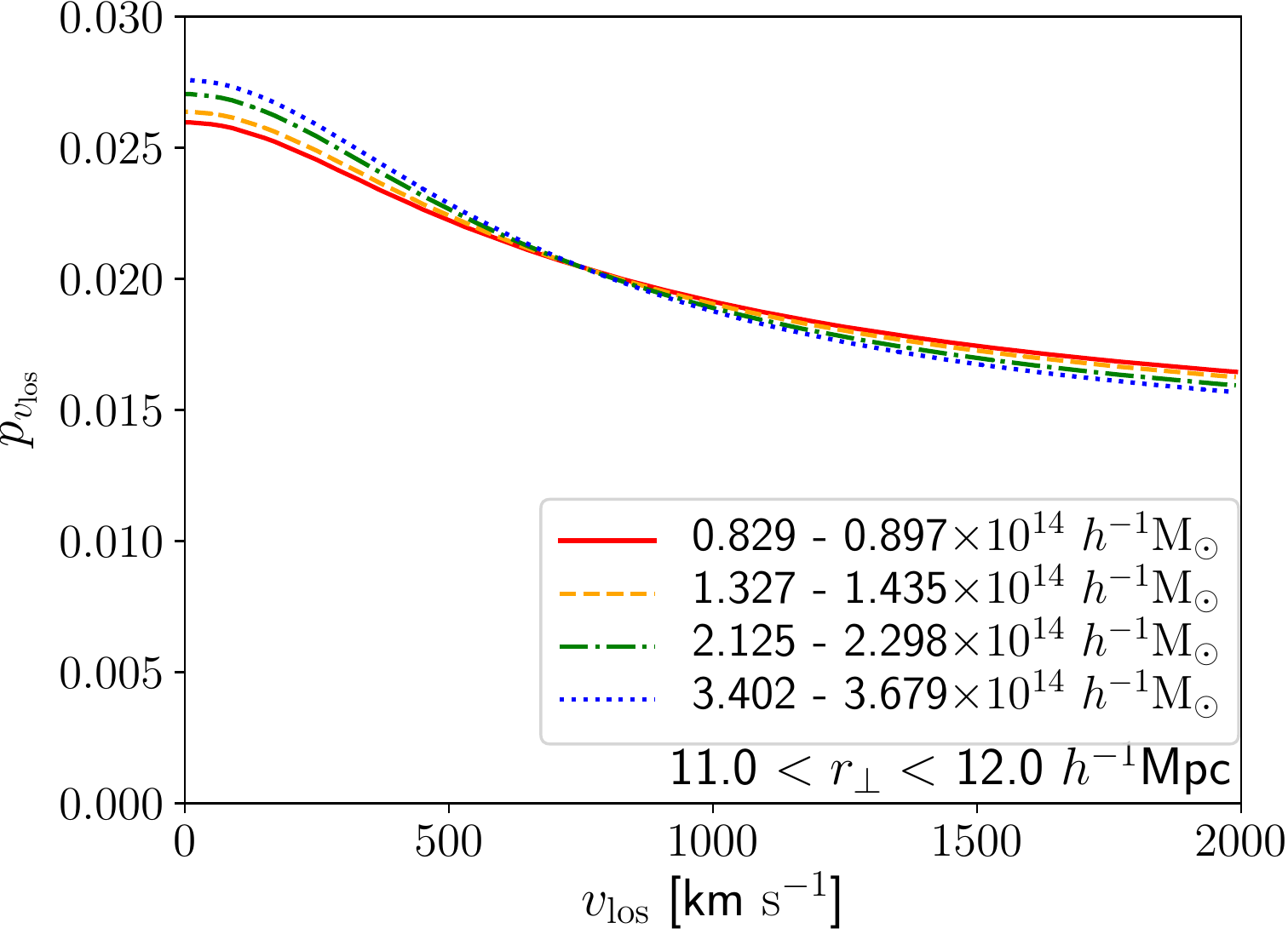}
       \end{minipage}
       \caption{Same as Fig. \ref{Fig7}, but for a fixed number density profile $n(r)$ (see text for details).}
    \label{Fig8}
\end{figure*}

Fig. \ref{Fig7} shows the cluster mass dependence of $p_{v_{\rm{los}}}$.
We can see a significant difference of $p_{v_{\rm{los}}}$ between different $M_{\rm{cl}}$ even at $r_{\perp}$ larger than $2  ~ h^{-1} {\rm{Mpc}} $.
The cluster mass dependence of $p_{v_{\rm{los}}}$ originates from the number density profile $n(r)$ and from the three-dimensional phase space distribution $p_{v}$.
Since our main interest is the cluster mass dependence purely coming from the latter, we check the cluster mass dependence of $p_{v_{\rm{los}}}$ for a fixed $n(r)$ but including the cluster mass dependence of $p_{v}$.
The result shown in Fig. \ref{Fig8} indicates that the cluster mass dependence is similar to that shown in Fig. \ref{Fig7}.
This confirms that the mass dependence from Fig. \ref{Fig7} mostly comes from $p_{v_{\rm los}}$ and we have a sufficiently large cluster mass dependence of $p_{v_{\rm los}}$ even if we fix $n(r)$ to the observed radial number density profile of galaxies.

\subsection{Origin of Cluster Mass Dependence of the PDF of $v_{\rm{los}}$}
\label{S3.3}
In Figs. \ref{Fig7} and \ref{Fig8}, we show that the cluster mass distribution of $p_{v_{\rm los}}$ is quite complicated despite the fact that the three-dimensional phase space distribution shows a rather simple cluster mass dependence (Fig. \ref{Fig4}).
Here we explore how such complicated cluster mass dependence appears.  
To do so, we check the contribution from different line-of-sight position $r_\parallel$ to $p_{v_{\rm los}}$.
Following equation (\ref{eq3-1}), we define $p_{v_{\rm{los}}} (v_{\rm{los}} , r_{\perp} , r_{\parallel})$ as
 \begin{equation}
 \label{eq3-5}
 \begin{split}
&p_{v_{\rm{los}}} (v_{\rm{los}} ,  r_{\perp} ,r_{\parallel}) = \\
& \frac{2}{N ( r_{\perp}) }  \int^{v_{{\rm{max}}}}_{ - v_{{\rm{max}}} } d v_{r}   \int^{v_{{\rm{max}}}}_{-v_{{\rm{max}}}} d v_{t}   n(r)  p_{v}(v_{r} , v_{t} , r)  \delta_{D} ( v_{\rm{los}} - v'_{\rm{los}} ) \ ,
\end{split}
\end{equation}
where $r$ is a function of $r_{\perp}$ and $r_{\parallel}$ as shown in equation (\ref{eq3-2}). 
\begin{figure}
    \includegraphics[width=\columnwidth]{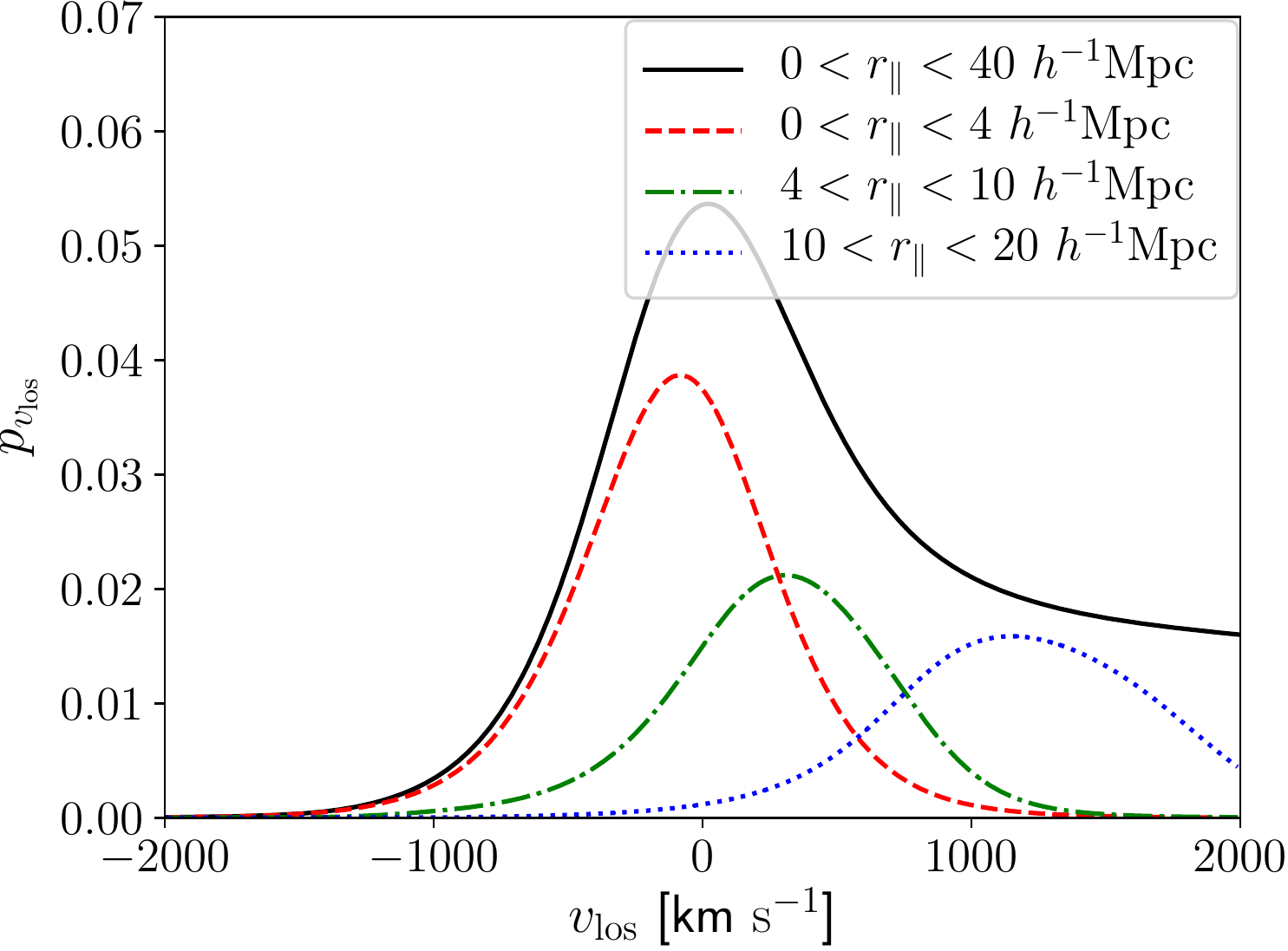}
 \caption{Comparison of $p_{v_{\rm{los}}} (v_{\rm{los}} ,  r_{\perp} , r_{\parallel})$ for four sets of parameters for $2.125 \times 10^{14} ~  h^{-1} M_{\odot} < M_{{\rm{cl}}} < 2.298 \times 10^{14} ~ h^{-1} M_{\odot} $.}
 \label{Fig9}
\end{figure}
%
In Fig. \ref{Fig9}, we show $p_{v_{\rm{los}}} (v_{\rm{los}} ,  r_{\perp} ,r_{\parallel})$ for four sets of parameters for a fiducial cluster mass.
We can see that high $r_{\parallel}$ segments contribute to the tail of $p_{v_{\rm{los}}}$ and low $r_{\parallel}$ segments contribute to the peak.
These are explained as follow.
At high $r_{\parallel}$ segments, the Hubble flow significantly increases $v_{\rm{los}}$, whereas at low $r_{\parallel}$ segments, the contribution of the Hubble flow is small so that we observe peculiar velocities directly.
We then check the dependence of each part of $p_{v_{\rm{los}}} (v_{\rm{los}} , r_{\perp} ,r_{\parallel})$ on cluster masses, focusing on $p_{v_{\rm{los}}} (v_{\rm{los}} , r_{\perp} ,r_{\parallel})$ at $2 ~ h^{-1} {\rm{Mpc}}   < r_{\perp} < 3 ~ h^{-1} {\rm{Mpc}} $.
\begin{figure*}
 \begin{minipage}{0.45\textwidth}
    \includegraphics[width=1.0\textwidth]{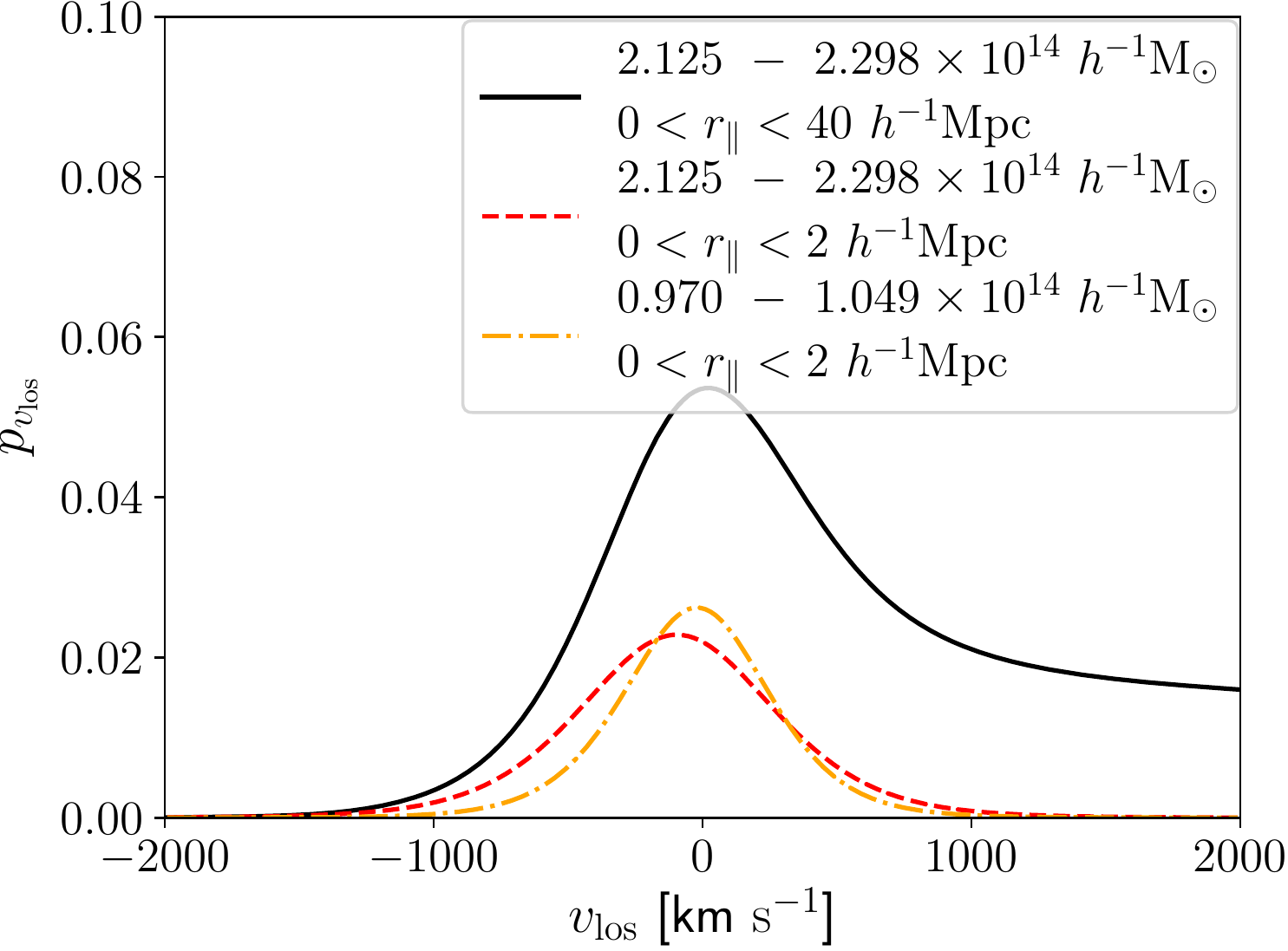}
           \end{minipage}
    \begin{minipage}{0.45\textwidth}
    \includegraphics[width=1.0\textwidth]{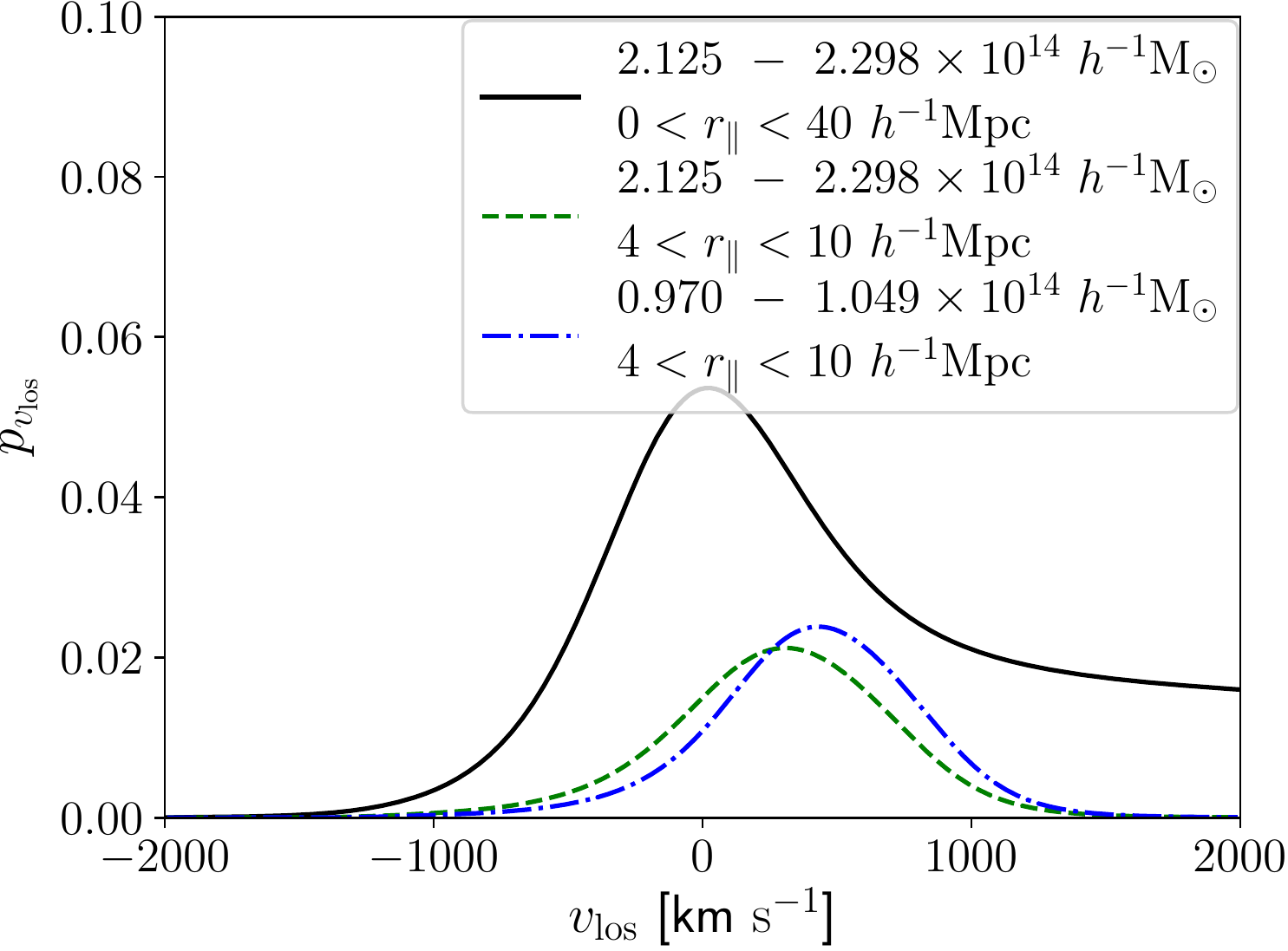}
       \end{minipage}
       \caption{$\it{Left}$: Comparison between $p_{v_{\rm{los}}} (v_{\rm{los}} ,  r_{\perp} ,r_{\parallel})$ at $0~ h^{-1} {\rm{Mpc}}  < r_{\parallel} < 40 ~ h^{-1} {\rm{Mpc}} $ ($\it{black \ solid}$), $0~ h^{-1} {\rm{Mpc}}  < r_{\parallel} < 2 ~ h^{-1} {\rm{Mpc}} $ ($\it{red \ dashed}$) for $2.125 \times 10^{14} ~ h^{-1} M_{\odot} < M_{{\rm{cl}}} < 2.298 \times 10^{14} ~ h^{-1} M_{\odot} $, and $0~ h^{-1} {\rm{Mpc}} < r_{\parallel} < 2 ~ h^{-1} {\rm{Mpc}} $ for $0.970 \times 10^{14} ~  h^{-1} M_{\odot} < M_{{\rm{cl}}} < 1.049 \times 10^{14} ~ h^{-1} M_{\odot} $ ($\it{orange \ dash-dotted}$).
$\it{Right}$: Comparison between $p_{v_{\rm{los}}} (v_{\rm{los}} ,  r_{\perp} ,r_{\parallel})$ at $0~ h^{-1} {\rm{Mpc}}  < r_{\parallel} < 40 ~ h^{-1} {\rm{Mpc}} $ ($\it{black \ solid}$), $4~ h^{-1} {\rm{Mpc}}  < r_{\parallel} < 10 ~ h^{-1} {\rm{Mpc}} $ ($\it{green \ dashed}$) for $2.125 \times 10^{14} ~ h^{-1} M_{\odot} < M_{{\rm{cl}}} < 2.298 \times 10^{14} ~ h^{-1} M_{\odot} $, and $4 ~ h^{-1} {\rm{Mpc}}   < r_{\parallel} < 10 ~ h^{-1} {\rm{Mpc}} $ for $0.970 \times 10^{14} ~ h^{-1} M_{\odot} < M_{{\rm{cl}}} < 1.049 \times 10^{14} ~ h^{-1} M_{\odot} $ ($\it{blue  \ dash-dotted}$).
Here we fix $r_{\perp}$ to $2 ~ h^{-1} {\rm{Mpc}}  < r_{\perp} < 3 ~ h^{-1} {\rm{Mpc}} $.}
    \label{Fig10}
\end{figure*}

First, we focus on a $r_{\parallel}$ segment with $r_{\parallel} \ll  r_{\perp}$.
In the left panel of Fig. \ref{Fig10}, we compare $p_{v_{\rm{los}}} (v_{\rm{los}} ,  r_{\perp} ,r_{\parallel})$ for the fiducial mass at $r_{\parallel} = [0 , 40] ~ h^{-1} {\rm{Mpc}} $ and $r_{\parallel} = [0 , 2] ~ h^{-1} {\rm{Mpc}} $.
We confirm that at low $r_{\parallel}$, $p_{v_{\rm{los}}} (v_{\rm{los}} ,  r_{\perp} ,r_{\parallel})$ contributes to the peak.
In the left panel of Fig. \ref{Fig10}, we also compare $p_{v_{\rm{los}}} (v_{\rm{los}} ,  r_{\perp} ,r_{\parallel})$ for fiducial mass and lower mass at $r_{\parallel} = [0 , 2] ~ h^{-1} {\rm{Mpc}} $. We can see that $p_{v_{\rm{los}}} (v_{\rm{los}} ,  r_{\perp} ,  r_{\parallel})$ for the higher cluster mass has a larger width than that for the lower cluster mass, because of larger $\sigma_{t} (r)$ for higher mass clusters as shown in Fig. \ref{Fig4}.
Note that at $r_{\parallel} \ll  r_{\perp}$ segments, $v_{r}$ does not contribute to $p_{v_{\rm{los}}}$ as shown in Section \ref{S3.1}.

Next we focus on a $r_{\parallel}$ segment with $r_{\parallel} >  r_{\perp}$.
In the right panel of Fig. \ref{Fig10}, we show a plot similar to the left panel of Fig. \ref{Fig10}, but we change $r_{\parallel}$ from $r_{\parallel} = [0 , 2] ~ h^{-1} {\rm{Mpc}} $ to $r_{\parallel} = [4 , 10] ~ h^{-1} {\rm{Mpc}} $.
In this segment, $v_{r}$, $v_{t}$, and the Hubble flow contributes to $v_{\rm{los}}$, and this segment contributes to the edge of the peak of $p_{v_{\rm{los}}}$.
While the effect of the Hubble flow is significant, $v_r$ also makes non-negligible contribution to the PDF.
Since clusters with higher masses have lower $v_r$ (see Fig. \ref{Fig4}), they shift $p_{v_{\rm{los}}} (v_{\rm{los}} ,  r_{\perp} ,r_{\parallel})$ to the lower mean $v_{\rm los}$ more significantly, which also results in the wider width of the PDF.

\section{Prospect for constraining cluster masses}
\label{S4.0}

\subsection{Accuracy of Mass Estimation}
\label{S4.01}
Our goal is to estimate masses of galaxy clusters by using $p_{v_{\rm{los}}}$ for a given $n(r)$.
We assume that we obtain observed $p_{v_{\rm{los}}}$ by stacking galaxies around many galaxy clusters and estimate the mean mass of galaxy clusters by using $p_{v_{\rm{los}}}$ in the range $2  ~h^{-1} {\rm{Mpc}} < r_{\perp} < 12 ~ h^{-1} {\rm{Mpc}} $.
We estimate the mean masses of galaxy clusters from observed $p_{v_{\rm{los}}}$ as follows.
First, we calculate $p_{v_{\rm{los}}}$ by using given $n(r)$ and $p_{v}$.
We compute $p_{v_{\rm{los}}}$ for different cluster masses, and calculate $\chi^2$ by comparing observed $p_{v_{\rm{los}}}$ with our model $p_{v_{\rm{los}}}$.
We determine the mean mass of galaxy clusters as the mass of our model $p_{v_{\rm{los}}}$ that results in the minimum $\chi^2$.

The $\chi^2$ can be described as
 \begin{equation}
 \label{eq4-0-1}
\chi^2 =
  \sum_{i}^{n_{\rm{bin}}} 
\frac{ ( \delta_{{\rm{stat}} , i} + \delta_{{\rm{ina}},i}  + \delta_{{M_{\rm{cl}} {\rm{dep}}} ,i} )^2 }
{ {\sigma_{i}}^{2} }  \ ,
\end{equation}
where $\delta_{{\rm{stat}} , i}$ is the difference between $p_{v_{\rm{los}}}$ directly measured from our $N$-body simulations and our model $p_{v_{\rm{los}}}$ purely caused by statistical fluctuations, $\delta_{{\rm{ina}} , i}$ is that caused by our model inaccuracy, $\sigma_{i}$ is the error of $p_{v_{\rm{los}}}$ directly measured from our $N$-body simulations (see Section \ref{S3.1}).
When we compare $p_{v_{\rm{los}}}$ directly measured from our $N$-body simulations with our model $p_{v_{\rm{los}}}$ with different cluster masses, there exists the difference arising from the difference between assumed and true cluster masses.
We denote this difference as $\delta_{{M_{\rm{cl}}}\rm{dep}}$.
As there is no correlation between $\delta_{{\rm{stat}} , i}$ and $\delta_{{\rm{ina}},i}$ or between $\delta_{{\rm{stat}} , i}$ and $\delta_{{M_{\rm{cl}} {\rm{dep}}} ,i}$, we rewrite equation (\ref{eq4-0-1}) as
\begin{equation}
\label{eq4-0-2}
\chi^2 =
\left \{  \sum_{i}^{n_{\rm{bin}}} 
\frac{ (\delta_{{\rm{ina}},i}  + \delta_{{M_{\rm{cl}} {\rm{dep}}} ,i} )^2 }
{ {\sigma_{i}}^{2} } \right \}  
+ {\chi_{\rm{stat}}}^2 \ ,
\end{equation}
where, 
\begin{equation}
\label{eq4-0-3}
{\chi_{\rm{stat}}}^2 \equiv
  \sum_{i}^{n_{\rm{bin}}} 
\frac{ \delta_{{\rm{stat}} , i}^2 }
{ {\sigma_{i}}^{2} }   \ .
\end{equation}
Note that ${\chi_{\rm{stat}}}^2$ must obey chi-square distribution with $k = n_{\rm{bin}}$.
Assuming that statistical errors derived from our $N$-body simulations are dominated by the Poisson errors, the dispersion $\sigma_{i}$ depends on the number of haloes in the $i$-th bin as
\begin{equation}
\label{eq4-0-4}
\sigma_{i} \sim \frac{ \sqrt{N_{{\rm{halo}} , i}} }{\sum_{i} N_{{\rm{halo}} , i} } \ ,
\end{equation}
where $N_{{\rm{halo}} , i}$ is the number of haloes at the $i$-th bin.
\begin{figure}
    \includegraphics[width=\columnwidth]{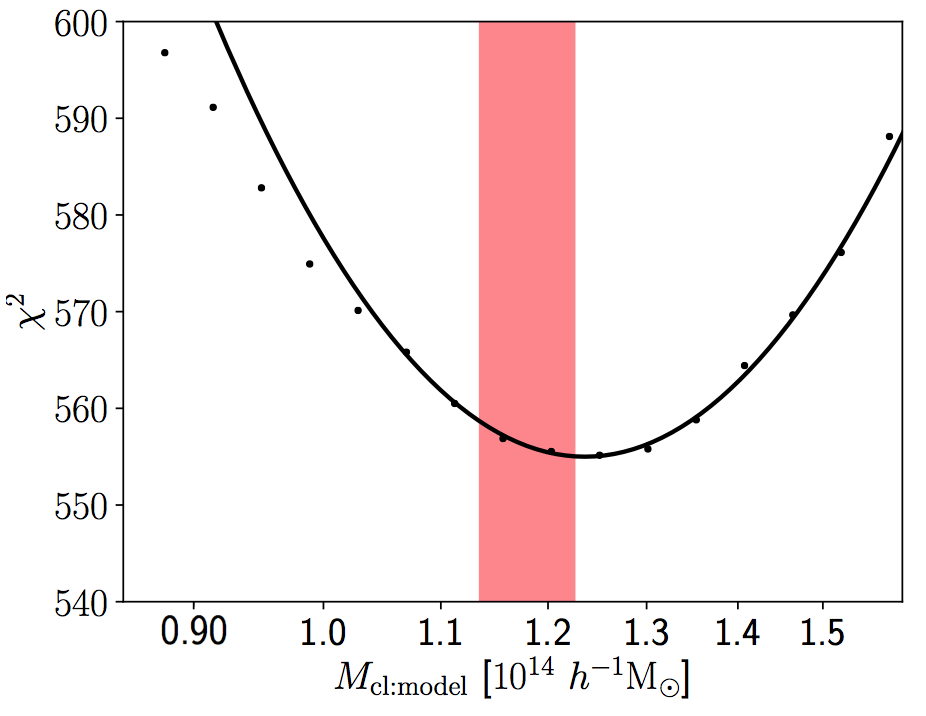}
 \caption{The distribution of $\chi^2 $ calculated from comparing $p_{v_{\rm{los}}}$ directly measured from our $N$-body simulations for a fixed cluster mass range with our model $p_{v_{\rm{los}}}$ for different cluster masses.
The shaded region shows the true cluster mass ($M_{{\rm{cl:true}}}$), and the horizontal axis shows the cluster mass of our model $p_{v_{\rm{los}}}$ ($M_{{\rm{cl:model}}}$).
The line shows the best fit line of equation (\ref{eq4-0-6}).
The total number of degree of freedom is 489.}
 \label{Fig11}
\end{figure}

Fig. \ref{Fig11} shows $\chi^2$ calculated by comparing $p_{v_{\rm{los}}}$ directly measured from our $N$-body simulations for a fixed cluster mass range with our model $p_{v_{\rm{los}}}$ for different cluster masses.
We fix $n(r)$ to the value that directly measured from our $N$-body simulations, and calculate $p_{v}$ for different cluster masses.
In total we use 431,925 haloes to derive $p_{v_{\rm los}}$ directly from our $N$-body simulations.

\begin{figure*}
 \begin{minipage}{0.45\textwidth}
    \includegraphics[width=1.0\textwidth]{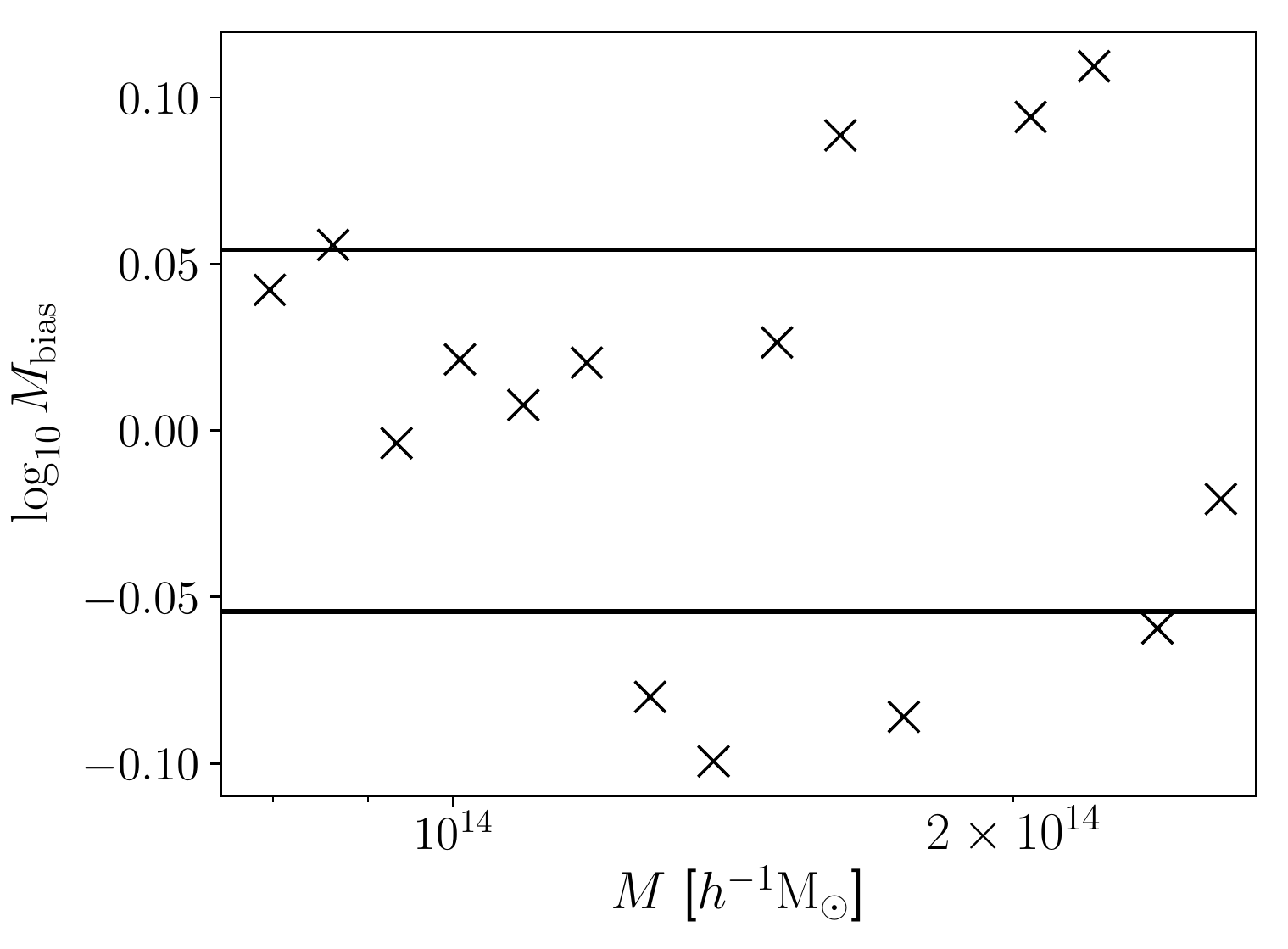}
           \end{minipage}
    \begin{minipage}{0.45\textwidth}
    \includegraphics[width=1.0\textwidth]{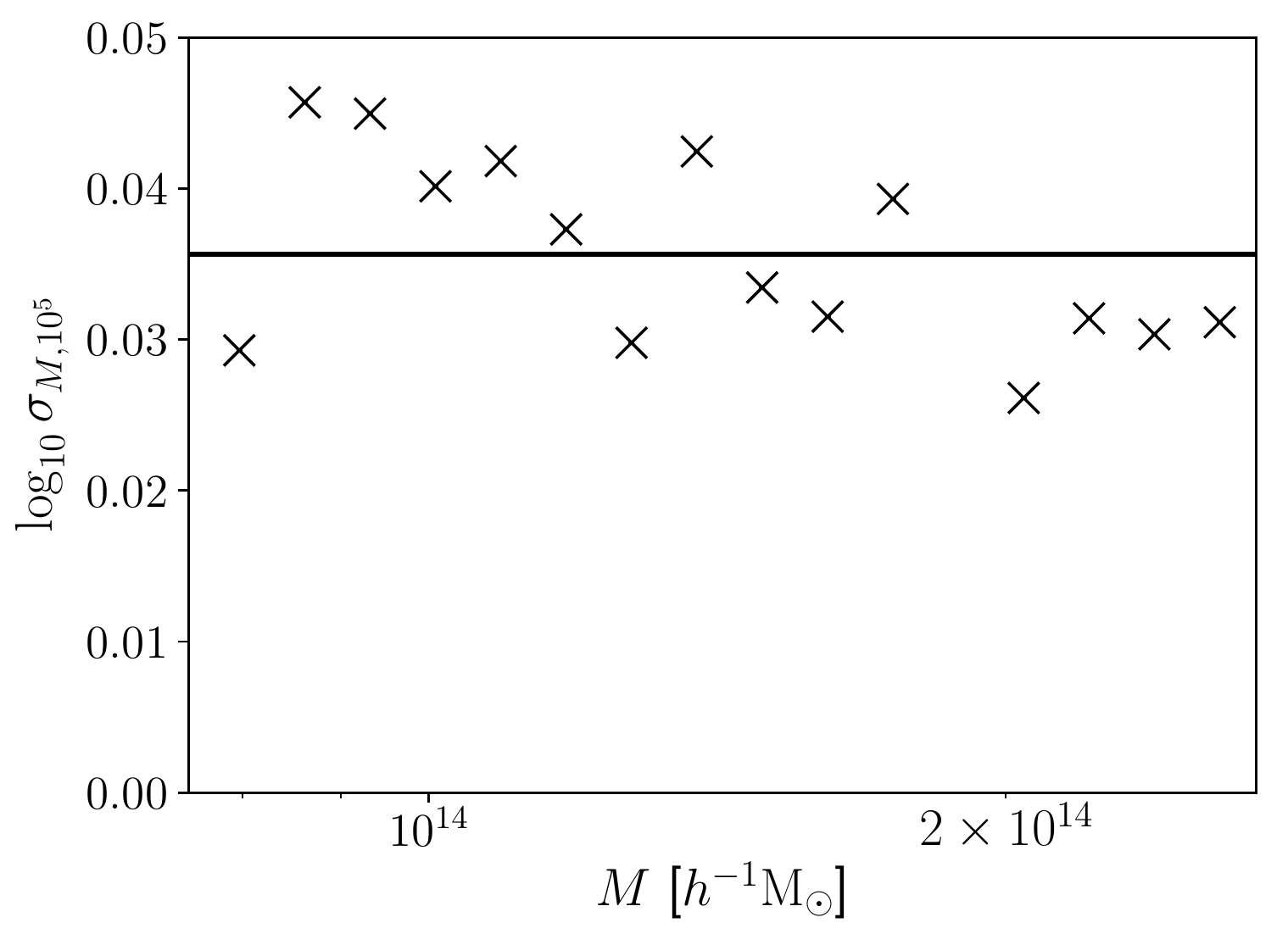}
       \end{minipage}
       \caption{$\it{Left}$: The systematic error of the cluster mass estimate due to an inaccuracy of our model, $ \log_{10}{M_{\rm{bias}}} $, as a function of input cluster masses.
The horizontal lines show the average of $| \log_{10}{M_{\rm{bias}}} |$.
$\it{Right}$: The statistical error on cluster mass estimate  when using $10^5$ haloes, $\log_{10}{\sigma_{M,10^{5}}} $, as a function of input cluster masses.
The horizontal line shows show the average of $\log_{10}{\sigma_{M,10^{5}}}$.}
    \label{Fig111}
\end{figure*}

Fig. \ref{Fig11} indicates that we can indeed constrain cluster masses with our approach, although there is an offset between input and best-fitting cluster masses, which originates from the inaccuracy of our model.
We fit the derived $\chi^{2}$ curve with the following function
\begin{equation}
\label{eq4-0-6}
\chi^{2} = 
\left \{\frac{ \log_{10}{ M_{{\rm{cl:true}}}} - \log_{10}{ M_{{\rm{cl:model}}}} + \log_{10}{M_{{\rm{bias}}}} }{\log_{10}{\sigma_{M}}} \right \}^2
 + C \ ,
\end{equation}
where $M_{\rm{bias}}$, $\sigma_{M}$, and $C$ are free parameters.
The best fit line is shown in Fig. \ref{Fig11}.
We obtain $M_{\rm{bias}} = 1.05$, $\sigma_{M}= 1. 05$, and $C = 555.00$ for Fig. \ref{Fig11}.
The parameter $M_{\rm{bias}}$ represents the systematic error of galaxy cluster mass estimation caused by the inaccuracy of our model, and $\sigma_{M}$ represents the $1\sigma$ statistical accuracy of cluster mass estimation.
As we see in equations (\ref{eq4-0-2}) and (\ref{eq4-0-4}), $\sigma_{M}$ depends on the number of haloes we use as
\begin{equation}
\label{eq4-0-7}
\begin{split}
\log_{10}{ \sigma_{M} }  \propto & \ \left \{  \sum_{i}^{n_{\rm{bin}}} 
\frac{ \delta_{{M_{\rm{cl}} {\rm{dep}}} ,i}^2 }
{ {\sigma_{i}}^{2} } \right \}^{-1/2}  \ \propto \frac{1}{ (N_{\rm{halo} })^{1/2} } \ ,
\end{split}
\end{equation}
where
\begin{equation}
\label{eq4-0-8}
N_{\rm{halo} } \equiv \sum_{i}^{n_{\rm{bin}}} N_{{\rm{halo} } , i} \ .
\end{equation}
We define $ \sigma_{M,10^{5}}$ the statistical accuracy of mass estimation for a fiducial number of $N_{\rm halo}$
\begin{equation}
\label{eq4-0-9}
\sigma_{M,10^{5}} \equiv {\sigma_{M}}^{\sqrt{N_{{\rm{halo} }} / {10^{5}}}} \ .
\end{equation}
Fig. \ref{Fig111} shows $ M_{\rm{bias}}$ and $ \sigma_{M,10^{5}}$ for different $M_{{\rm{cl:true}}}$.
We obtain $ M_{\rm{bias}}$ and $ \sigma_{M,10^{5}}$ by the same way as done in Fig. \ref{Fig11}.
We can see that $ M_{\rm{bias}}$ and $ \sigma_{M,10^{5}}$ have little dependence on cluster masses.
The average of $| \log_{10}{M_{\rm{bias}}} |$ is $0.05$, which corresponds to the systematic error of cluster mass estimation caused by the inaccuracy of our model of $13.33 \%$.
The average of $\log_{10}  \sigma_{M,10^{5}} $ is $0.04$, which corresponds to the statistical error with $N_{{\rm{halo} } } = 10^5$ of $8.55 \%$.

\begin{figure}
  \includegraphics[width=\columnwidth]{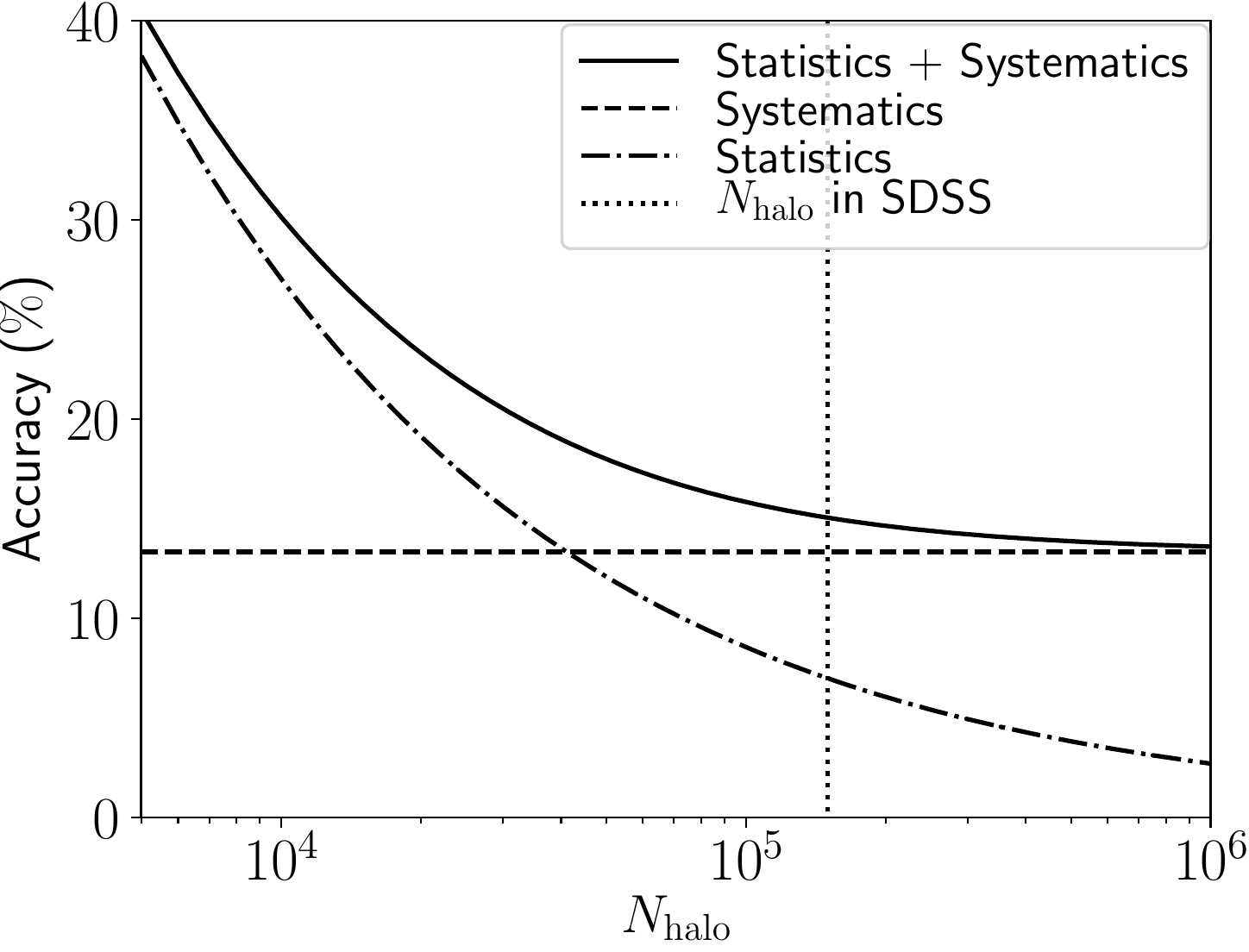}
  
    \caption{The accuracy of the mean mass estimation of clusters as a function of the total number of haloes, $N_{{\rm{halo}}}$. 
The dashed line shows the systematic error of the cluster mass estimation caused by the inaccuracy of our model, the dash-dotted line shows the the statistical error as a function of $N_{{\rm{halo}}}$, and the solid line shows the quadrature sum of the systematic and statistical errors.
The vertical dotted line corresponds to the number of SDSS/BOSS spectroscopic galaxies we can use for the analysis.}
 \label{Fig112}
\end{figure}

Our result described above allows us to estimate the accuracy of cluster mass estimates as a function of the number of haloes, $N_{\rm halo}$.
Fig. \ref{Fig112} shows the dependence of the accuracy of the mean mass estimation on $N_{\rm halo}$ both from the statistical and systematic errors.
We consider the systematic error of cluster mass estimation caused by the inaccuracy of our model.
This result can be used to derive expected errors of average cluster masses from various observations.
For instance, the number of spectroscopic galaxies at $2 ~ h^{-1} {\rm{Mpc}} < r_{\perp} < 12 ~ h^{-1} {\rm{Mpc}}$ for clusters with richness $20 < N < 60$ at redshift $0.1 < z < 0.4$ observed in SDSS/BOSS \citep{Dawson2013} is about $1.5 \times 10^{5}$.
Here we use richness $N$ estimated by CAMIRA \citep{Oguri2014}.
We can determine the mean mass of galaxy clusters using spectroscopic galaxies in SDSS/BOSS from the PDF of $v_{{\rm{los}}}$ at $2 ~ h^{-1} {\rm{Mpc}} < r_{\perp} < 12 ~ h^{-1} {\rm{Mpc}}$ with an accuracy of $ 14.5 \%$, where the error include both the statistical and statistical error.
We can reduce the systematic error by improving the accuracy of the model of the phase space distribution. Suppose that we construct a sufficiently accurate model, we can reduce the error down to 5.7\%, which corresponds to the pure statistical error estimated above.


\subsection{Other Possible Systematics}
\label{S4.1}

\subsubsection{Halo Mass Dependence}
\label{S4.1.1}
In this paper, we use haloes with masses $M_{{\rm{halo}}} > 1 \times 10^{11} ~ h^{-1} M_{\odot} $ as haloes. 
However, in observations, we may not be able to accurately estimate halo masses of spectroscopic galaxies that we use for the stacked phase space distribution analysis.
Any mismatch between assumed and true halo masses can therefore become a source of additional systematic error.

In order to evaluate the possible systematic error from the mismatch of halo masses, we calculate $ M_{\rm{bias}}$ and $ \sigma_{M}$ for haloes adopting larger masses of  $M_{{\rm{halo}}} > 2 \times 10^{11} ~ h^{-1} M_{\odot} $.
We use $p_{v}$ calculated for $M_{{\rm{halo}}} > 1 \times 10^{11} ~ h^{-1} M_{\odot} $ and $n(r)$ directly measured in our $N$-body simulations for $M_{{\rm{halo}}} > 2 \times 10^{11} ~ h^{-1} M_{\odot} $.
We obtain $ M_{\rm{bias}}$ and $ \sigma_{M}$ by the same way as done in Fig. \ref{Fig11}.
We find that the average of $|\log_{10} M_{\rm{bias}}|$ is $0.06$, which corresponds to the systematic error of cluster mass estimation caused by the inaccuracy of our model of $14.34 \%$.
The average of $\log_{10} \sigma_{M, 10^{5}}$ is $0.03$, which corresponds that the statistical error with $N_{{\rm{halo} } } = 10^{5}$ is $7.94 \%$.
Since these values are quite similar to those derived in Section \ref{S4.01}, we conclude that the systematic error from the inaccuracy of the masses of haloes is small, at least compared with  that caused by the inaccuracy of our model.

\subsubsection{Miscentreing}
\label{S4.1.2}
In observation, we use BCGs defined by cluster finding methods as centres of clusters, whereas in our $N$-body simulations, we determine the most bounded dark matter particles as centres of clusters according to $\tt{Rockstar}$ \citep{Behroozi2013B}.
These BCGs may not correspond to the centres defined in our $N$-body simulations.
This miscentreing causes systematic errors. 

In \citet{Rykoff2016}, they compare the sky positions of the BCGs of the clusters observed by SDSS \citep{SDSS} defined in their cluster finding method \citep{Rykoff2014} with the sky positions of the X-ray peak of the clusters observed by $Chandra$ and $XMM$ X-ray. In \citet{Rykoff2016}, they also compare the sky positions of the BCGs with the sky position of the peak of SZ signals observed from the South Pole Telescope SZ cluster survey \citep{Bleem2015}.
While we want to estimate distances between centres of clusters defined by $\tt{Rockstar}$ and BCGs defined by their cluster finding method, it is difficult to find centres of clusters defined by $\tt{Rockstar}$ in observations.
Hence, they use peaks of X-ray and SZ signals as proxy to true cluster centres to estimate the frequency of miscentreing by their cluster finding method.

In \citet{Rykoff2016}, they model the PDF of the miscentreing effect as
\begin{equation}
 \label{eq4-1-3-1}
p (x)= \frac{\rho_{0}}{\sigma_{0} \sqrt{2 \pi}} \exp{ \frac{- x^2}{ 2 \sigma_{0}^2}} + \frac{(1 - \rho_{0}) x}{ \sigma_{1}^2 } \exp{ \frac{- x^2}{ 2 \sigma_{1}^2} },
\end{equation}
where $\rho_{0}$, $\sigma_{0}$, and $\sigma_{1}$ are free parameters, $x = r / R_{\lambda}$, $r$ is the projected radius of the BCGs from peaks of X-ray and SZ signals, and $R_{\lambda}$ is the projected radius of the galaxy clusters with richness $\lambda$ (see e.g., \citealt{Rykoff2014}; \citealt{Oguri2014}).
The first term of right hand side corresponds to the BCGs-X-ray (SZ) centre offset, and the second term represents systematics failures in identifying the correct BCGs with their cluster finding method.
They obtained $\rho_{0} = 0.78^{+0.11}_{-0.11}$ and $\sigma_{1} = 0.31^{+0.09}_{-0.05}$ by comparing the offset distribution derived from the observations mentioned above with equation (\ref{eq4-1-3-1}).
They marginalise over the parameter $\sigma_{0}$ since it is not relevant to the overall fraction of misidentified BCGs.

The miscentreing affects $p_{v_{\rm{los}}}$ in two aspects. First, we mis-measure redshifts of BCGs.
This causes systematic errors to $v_{\rm{los}}$.
Second, we mis-measure the projected radii to member galaxies.
The first aspect distorts $p_{v_{\rm{los}}}$ as
\begin{equation}
 \label{eq4-1-3-3}
p_{v_{{\rm{los}}} : {\rm{obs}}}  (v_{{\rm{los}}})= \int^{\infty}_{- \infty} d v'_{{\rm{los}}} p_{v_{{\rm{los}}} : {\rm{true}}} (v'_{{\rm{los}}}) E( v_{{\rm{los}}} - v'_{{\rm{los}}}) ,
\end{equation}
where $p_{v_{{\rm{los}}} : {\rm{obs}}}$ and $p_{v_{{\rm{los}}} : {\rm{true}}} $ are the observed and the true PDFs of the line-of-sight velocity, $E$ is the effect of the mis-measure redshifts of BCGs, which is described as
\begin{equation}
 \label{eq4-1-3-3}
E (x)= \frac{1 - \rho_{0}}{\sigma_{{\rm{BCG}}} \sqrt{2 \pi}} \exp{\frac{-x^2}{2 \sigma_{{\rm{BCG}}}^2}} + \rho_{0} \delta (x),
\end{equation}
where $\sigma_{\rm{BCG}}$ is the typical line-of-sight velocity between correct BCGs and galaxies misidentified as BCGs.
Estimating values of $\sigma_{\rm{BCG}}$ and discussions of the second aspect are left for future work.

\section{Summary and Discussion}
\label{SD}
We have constructed a model of the three-dimensional phase space distribution of haloes surrounding galaxy clusters up to 50 $h^{-1}$Mpc from cluster centres based on the stacked phase space distribution from $N$-body simulations.
We have adopted a two component model with flexible radial and tangential velocity distributions for each component, in order to accurately model the complex phase space distribution seen in the $N$-body simulations.
By projecting the three-dimensional phase space distribution along the line-of-sight with the effect of the Hubble flow, we have obtained a model of the project phase space distribution as a function of the line-of-sight velocity, $p_{\rm v_{\rm los}}$. 
We have derived $p_{v_{\rm los}}$ as a smooth function of both projected transverse distance from cluster centres as well as cluster masses.
We have shown that $p_{v_{\rm los}}$ shows a complicated dependence on cluster masses due to the interplay between the infall velocity and the Hubble flow.
Even after matching the number density profile of haloes, $p_{v_{\rm los}}$ at $r_{\perp}>2h^{-1}$Mpc shows a significant cluster mass dependence, indicating that we can constrain cluster masses from the stacked phase space distribution at large radii.

Using our model, we have estimated the accuracy of the cluster mass estimation.
We have found that, when using $1.5 \times 10^{5}$ spectroscopic galaxies, we can constrain the mean cluster mass from  $p_{v_{\rm{los}}}$ in the range $2  ~ {\rm{Mpc}} / {\it{h}} < r_{\perp} < 12 ~ {\rm{Mpc}} / {\it{h}} $ with an accuracy of $14.5 \%$, where the error includes both the statistical error and the systematic error from an inaccuracy of our model.
By improving the accuracy of our model, the error on the mean mass can be improved down to 5.7\%, which suggests that the method we propose in the paper has a great potential to constrain cluster masses.
We can improve our model by $(1)$ using $N$-body simulations with more realizations and obtain parameters of the three-dimensional phase space distribution more accurately, and $(2)$ adopting better functional form of the three-dimensional phase space distribution.
Nevertheless, our work represents the first step toward constraining cluster masses from the stacked phase space distribution at large radii, and quantifies its potential for future applications. 

We have shown that our new method can constrain cluster masses with an accuracy comparable to other methods, e.g., the weak lensing effect, $X$-ray observation, and dynamical mass measurements from galaxy motions inside clusters.
Given a rapid progress of wide-field spectroscopic surveys, our new approach will grow its importance, particularly at high redshifts where weak lensing measurements of cluster masses become more difficult.
Therefore it is interesting to check the potential of our method in future spectroscopic surveys such as Subaru Prime Focus Spectrograph (PFS; \citealt{Takada2014}) and Dark Energy Spectroscopic Instrument (DESI; \citealt{DESI}).

\section*{Acknowledgements}
We thank Yasushi Suto, Kazuhisa Mituda, Masami Ouchi, and Taizo Okabe for useful discussions, and Ying Zu for useful correspondence.
This work was supported in part by World Premier International Research Center Initiative (WPI Initiative), MEXT, Japan, and JSPS KAKENHI Grant Number JP15H05892 and JP18K03693.
TN is supported in part by JSPS KAKENHI Grant Number 17K14273, and JST CREST Grant Number JPMJCR1414.
Numerical simulations were carried out on Cray XC50 at the Center for Computational Astrophysics, National Astronomical Observatory of Japan.



\bibliography{reference}

\bibliographystyle{mnras}



\appendix
\section{Parameters}
In Tables \ref{T1} and \ref{T2}, we list the parameters of the three-dimensional phase space distribution we adopted in this paper.
We use these parameters in equations (\ref{eq2-12}), (\ref{eq2-13}), and (\ref{eq2-17}).

\begin{table}
\begin{center}
 \caption{Sets of parameters shown in equations (\ref{eq2-12}) and  (\ref{eq2-13})}
 \begin{tabular}{| c | c c c c c c |}
  \hline
    & $A_{,1}$ & $A_{,2}$ & $A_{,3}$ & $A_{,4}$ & $A_{,5}$ & $A_{,6}$\\  \hline
 $\delta_{r}$ & $1.936 $ & $0.3926 $ &  $216.1 $ & $130.4 $ & $1.671$ & $0.7441 $  \\  
  $\delta_{t}$ & $2.925 $ & $ 0.6693 $ &  $1.878 \times 10^{-2}$ & $ 0 $ & $ 0 $ & $1.270$  \\ 
  $\gamma_{r}$ & $ 18.20 $ & $ 0.9994 $ &  $7.650$ & $ 2.687 $ & $ -18.14 $ & $0.2512 $ \\  
  \hline
 \end{tabular}
         \label{T1} 
              \end{center}
             \end{table}
\begin{table}
\begin{center}
 \caption{Sets of parameters shown in equation (\ref{eq2-17})}
 \begin{tabular}{| c | c c c |}
  \hline
    & $B_{,1}$ & $B_{,2}$ & $B_{,3}$ \\  \hline
$\alpha , 1$ & $6.663 \times 10^{-1} $ & $2.256 \times 10^{-3}$ &  $ 0 $  \\   
$\alpha , 2$ & $7.089 \times 10^{-1} $ & $2.927 \times 10^{-1}$ &  $ 0 $  \\    
$\xi_{r} , 1$ & $ -95.33 $ & $7.023 \times 10^{-1}$ &  $ 164.4 $  \\   
 $\xi_{r} , 2$ & $ -44.88 $ & $6.433 \times 10^{-3}$ &  $ 4.544 $  \\   
  $\xi_{r} , 3$ & $ 1.613 $ & $2.892 \times 10^{-1}$ &  $ 1.561 $  \\   
  $\xi_{r} , 4$ & $ 6660 $ & $ -9.865 \times 10^{-1}$ &  $ -49.99 $  \\   
   $\lambda_{r} , 1$ & $ 5.387 \times 10^{-4} $ & $ 3.495$ &  $ 333.5 $  \\   
   $\lambda_{r} , 2$ & $ 4.542 \times 10^{-6} $ & $ 2.663$ &  $ 3.369  \times 10^{-1}$  \\ 
    $\lambda_{r} , 3$ & $ 8.462 \times 10^{-1} $ & $ 2.647 \times 10^{-1}$ &  $ 16.44$  \\ 
    $\lambda_{r} , 4$ & $ 64.92 $ & $ 2.247 \times 10^{-1}$ &  $ 88.94$  \\ 
    $\mu_{r} , 1$ & $ 1.235 \times 10^{4} $ & $ 2.082 \times 10^{-2}$ &  $ -1.287 \times 10^{4} $  \\ 
    $\mu_{r} , 2$ & $ 5.223 \times 10^{-4} $ & $ 1.428 \times 10^{-1}$ &  $ -3.464 \times 10^{-4} $  \\ 
    $\mu_{r} , 3$ & $ 0 $ & $ 0 $ &  $ 0 $  \\ 
    $\mu_{r} , 4$ & $ 1.745 \times 10^{6}$ & $ 5.271 \times 10^{-2} $ &  $ -1.276 \times 10^{6} $  \\ 
    $\sigma_{r} , 1$ & $ 5.027 \times 10^{5} $ & $ -2.363 $ &  $ 537.5 $  \\ 
     $\sigma_{r} , 2$ & $ 36.11 $ & $ -8.217 \times 10^{-1} $ &  $ -1.366 $  \\ 
      $\sigma_{r} , 3$ & $ -5.911 \times 10^{-1} $ & $ 1.471 $ &  $ 176.6 $  \\ 
       $\sigma_{r} , 4$ & $ 11.82 \times 10^{-1} $ & $ 1.471 $ &  $ 353.2 $  \\
       $\lambda_{t} , 1$ & $ 2609  $ & $ 1.441 \times 10^{-1}$ &  $ -2453 \times  $  \\ 
         $\lambda_{t} , 2$ & $ 5.823  $ & $ -1.401 $ &  $ 6.106 \times 10^{-1} $  \\ 
          $\lambda_{t} , 3$ & $ -7.840  $ & $ 9.630 \times 10^{-2} $ &  $ 19.25  $  \\ 
           $\lambda_{t} , 4$ & $ 436.4  $ & $ 1.105 \times 10^{-1} $ &  $ -268.5  $  \\ 
            $\sigma_{t} , 1$ & $ 76.87  $ & $ 8.602 \times10^{-1} $ &  $ -254.4 $  \\
              $\sigma_{t} , 2$ & $ -9381  $ & $ -3.995 $ &  $ 2.177 $  \\
              $\sigma_{t} , 3$ & $ 4841  $ & $ -1.680 $ &  $ 2.607 $  \\ 
              $\sigma_{t} , 4$ & $ 201.4  $ & $ 6.859 \times 10^{-1} $ &  $ -600.3 $  \\

  \hline
 \end{tabular}
         \label{T2} 
              \end{center}
             \end{table}


\bsp	
\label{lastpage}
\end{document}